\documentclass[useAMS]{mn2e}

\usepackage{multirow}
\usepackage{bigdelim}
\usepackage{longtable}
\usepackage{graphicx}
\usepackage{times}

\def\S{SVS13 }
\def\SA{SVS13-A }

\def \form13{H$_2^{13}$CO }

\def \formd2{D$_2$CO }

\def \met13{$^{13}$CH$_3$OH }

\def \metan2d{CHD$_2$OH }
\def \metan3d{CH$_3$OD }

\usepackage{fixltx2e}
\title[]{Decrease of the organic deuteration during the evolution of Sun-like protostars: the case of SVS13-A}

\author[E. Bianchi et al.]{E. Bianchi$^{1,2}$\thanks{E-mail:
ebianchi@arcetri.astro.it} , C. Codella$^{1}$, C. Ceccarelli$^{3,4,1}$, F. Fontani$^{1}$, L. Testi$^{5,1}$,  R. Bachiller$^{6}$, 
 \newauthor B. Lefloch$^{3,4}$, L. Podio$^{1}$, V. Taquet$^{7}$
\\
$^{1}$ INAF-Osservatorio Astrofisico di Arcetri, L.go E. Fermi 5, Firenze, 50125, Italy \\
$^{2}$ Dipartimento di Fisica e Astronomia, Universit\`a degli Studi di Firenze, Italy \\
$^{3}$ Univ. Grenoble Alpes, IPAG, F-38000 Grenoble, France \\
$^{4}$ CNRS, IPAG, F-38000 Grenoble, France \\
$^{5}$ ESO, Karl Schwarzschild str. 2, D--85748 Garching bei Muenchen, Germany\\
$^{6}$ IGN, Observatorio Astron\'omico Nacional, Calle Alfonso XIII, 28004 Madrid, Spain\\
$^{7}$ Leiden Observatory, Leiden University, PO Box 9513, 2300--RA, Leiden, The Netherlands
}

\begin{document}

\date{Accepted date. Received date; in original form date}

\pagerange{\pageref{firstpage}--\pageref{lastpage}} \pubyear{2017}

\maketitle

\label{firstpage}

\begin{abstract}

We present the results of formaldehyde and methanol deuteration measurements towards the Class I low-mass protostar SVS13-A, in the framework of the IRAM 30-m ASAI (Astrochemical Surveys At IRAM) project. We detected emission lines of formaldehyde, methanol, and their deuterated forms (HDCO, D$_2$CO, CHD$_{2}$OH, CH$_{3}$OD) with $E_{\rm up}$ up to 276 K. The formaldehyde analysis indicates $T_{\rm kin} \sim$ 15 -- 30 K, $n_{\rm H_2} \geq$ 10$^{6}$ cm$^{-3}$, and a size of about 1200 AU suggesting an origin in the protostellar envelope. For methanol we find two components: (i) a high temperature ($T_{\rm kin} \sim$ 80 K) and very dense ($>$ 10$^{8}$ cm$^{-3}$) gas from a hot corino (radius $\simeq$  35 AU), and (ii) a colder ($T_{\rm kin} \leq$ 70 K) and more extended (radius $\simeq$ 350 AU) region. 

The deuterium fractionation is 9 $\times$ 10$^{-2}$ for HDCO, 4 $\times$ 10$^{-3}$ for D$_{2}$CO, and 2 -- 7 $\times$ 10$^{-3}$ for CH$_2$DOH, up to two orders of magnitude lower than the values measured in Class 0 sources. 
We derive also formaldehyde deuteration in the outflow: 4 $\times$ 10$^{-3}$, in agreement with what found in the L1157--B1 protostellar shock.
Finally, we estimate [CH$_{2}$DOH]/[CH$_{3}$OD] $\simeq$ 2. The decrease of deuteration in the Class I source SVS13-A with respect to Class 0 sources can be explained by gas-phase processes. Alternatively, a lower deuteration could be the effect of a gradual collapse of less deuterated external shells of the protostellar evelope. The present measurements fill in the gap between prestellar cores and protoplanetary disks in the context of organics deuteration measurements.

\end{abstract}


\section{Introduction}

Deuterium fractionation is the process that enriches the amount
of deuterium with respect to hydrogen in molecules.
While the D/H elementary abundance ratio is $\sim$ 1.6 $\times$ 10$^{-5}$ (Linsky 2007),
for molecules this ratio can be definitely higher and can be a precious
tool to understand the chemical evolution of interstellar gas (see e.g. Ceccarelli et al. 2014, and references therein).
In particular, during the process leading to the formation of a Sun-like
star, large deuteration of formaldehyde and methanol is observed in
cold and dense prestellar cores (e.g. Bacmann et al. 2003, Caselli \& Ceccarelli 2012, and
references therein).
Formaldehyde can be formed through gas phase chemistry in prestellar cores (Roberts \& Millar 2000b). The picture is different for formaldehyde as well as for methanol around protostars which are mostly formed via active grain surface chemistry 
(e.g. Tielens 1983). Deuterated H$_2$CO and CH$_3$OH are then stored
in the grain mantles to be eventually released into the gas phase once the protostar is formed 
and the grain mantles are heated and successively evaporated
(e.g. Ceccarelli et al. 1998, 2007; Parise et al. 2002, 2004, 2006) or sputtered by protostellar shocks
(Codella et al. 2012; Fontani et al. 2014).
As a consequence, D/H can be used as fossil record of
the physical conditions at the moment of the icy water and organics formation (e.g. Taquet et al.
2012, 2013, 2014).

While deuterated molecules have been detected towards 
the early stages of the Sun-like star formation 
(i.e. prestellar cores and Class 0 objects) as well as in the Solar System (see e.g. Ceccarelli et al. 2014, and references therein),
 no clear detection has been obtained for 
intermediate evolutionary phases (Class I and II objects).
 A handful of measurements of deuterium fractionation in
 Class I sources exists (i.e. Roberts \& Millar 2007)
but they refer only to few transitions 
sampling large regions (up to 58$\arcsec$) well beyond
the protostellar system.
In addition, Loinard et al. (2002) reported measurements of double deuterated formaldehyde in star-forming 
regions with both the SEST (Swedish ESO Submillimeter telescope) and IRAM single
dishes suggesting a decrease with the evolutionary stage.
Watanabe et al. (2012) reported the deuterium fractionation measurements toward 
R CrA IRS7B, a low-mass protostar in the Class 0/I transitional stage. 
They detected H$_{2}$CO, measuring a lower D/H ($\sim$ 0.05) compared to deuteration measured in Class 0 objects.
However, in this case the low deuterium fractionation ratios do not directly suggest an evolutionary trend. The altered chemical composition of the envelope of R CrA IRS7B can be a result of the heating of the protostar parent core by the external UV radiation from the nearby Herbig Ae star R CrA.

Systematic observations of D/H in Class I objects 
are therefore required to understand how the deuterium fractionation
evolves from prestellar cores to protoplanetary
disks.
In this context we present a study of formaldehyde and methanol 
deuteration towards
the Class I low-mass protostar SVS13-A.

\subsection{The SVS13 star forming region}
The SVS13 star forming region is located in the NGC1333 cloud in
Perseus at a distance of 235 pc (Hirota et al. 2008). 
It is associated with a Young Stellar Objects (YSOs) cluster,
dominated, in the millimeter by
two objects, labelled A, and B, respectively, separated by $\sim$ 15$\arcsec$.
(see e.g. Chini et al. 1997; Bachiller et al. 1998; Looney et al.
2000; Chen et al. 2009; Tobin et al. 2016, and references therein).
Interestingly, SVS13-A and SVS13-B are associated with two different
evolutionary stages. On the one hand, SVS13-B is a Class 0 protostar
with $L_{\rm bol}$ $\simeq$ 1.0 $L_{\rm sun}$ (e.g. Tobin et al. 2016)
driving a well collimated SiO jet (Bachiller et al. 1998).
On the other hand, SVS13-A is definitely more luminous ($\simeq$ 32.5
$L_{\rm sun}$, Tobin et al. 2016) and is associated
with an extended outflow ($>$ 0.07 pc, Lefloch et al. 1998, Codella et
al. 1999) as well as with the well-known chain of
Herbig-Haro (HH) objects 7-11 (Reipurth et al. 1993). In addition,
SVS13-A has a low $L_{\rm submm}$/$L_{\rm bol}$ ratio ($\sim$ 0.8\%)
and a high bolometric temperature ($T_{\rm bol}$ $\sim$ 188 K, Tobin
et al. 2016).
Thus, although still deeply embedded in a large scale envelope
(Lefloch et al. 1998), SVS13-A is considered a more evolved
protostar, already entered in the Class I stage. 
For all these reasons, SVS13-A is an almost unique laboratory
to investigate how deuteration change from the Class 0 to the Class I phases.
In Sect. 2 the IRAM 30-m observations are described, in Sect. 3
we report the results, while in Sect. 4 we develop the analysis of the data;
Sect. 5 is for the conclusions.

\section{Observations}
The observations of \SA were carried out 
with IRAM 30-m telescope near Pico Veleta (Spain), in the framework of
the Astrochemical Surveys At IRAM\footnote{www.oan.es/asai} (ASAI)
Large Program.
 The data consist of an unbiased spectral survey
acquired during several runs between 2012 and 2014, using the
broad-band EMIR receivers.
In particular the observed bands are at 3 mm (80--116 GHz), 2 mm
(129--173 GHz), and 1.3 mm (200--276 GHz).
The observations were acquired in wobbler switching mode, with a throw
of 180$\arcsec$ towards the coordinates
of SVS13-A, namely
$\alpha_{\rm  J2000}$ = 03$^{\rm h}$ 29$^{\rm m}$ 10$\fs$42,
$\delta_{\rm J2000}$ = +31$\degr$ 16$\arcmin$ 0$\farcs$3
The pointing was checked by observing nearby planets or continuum
sources and was found to be accurate to within 2$\arcsec$--3$\arcsec$.
The telescope HPBWs range between $\simeq$ 9$\arcsec$ at 276 GHz to $\simeq$ 30$\arcsec$ at 80 GHz.
The data reduction was performed using the
GILDAS--CLASS\footnote{http://www.iram.fr/IRAMFR/GILDAS} package.
Calibration uncertainties are stimated to be $\simeq$ 10\% at 3 mm and
$\sim$ 20\% at lower wavelengths.
Note that some lines (see Sect. 3) observed at 2 mm and 3 mm (i.e. with
a HPBW $\geq$ 20$\arcsec$) are affected by emission at OFF
position observed in wobbler mode.
Line intensities have been converted from antenna temperature to main
beam temperature ($T_{\rm MB}$), using the main
beam efficiencies reported in the IRAM 30-m
website\footnote{http://www.iram.es/IRAMES/mainWiki/Iram30mEfficiencies}.

\section{Results}\label{sec:Results}

\subsection{Line identification}\label{sec:line_id}

Line identification has been performed using a package developed at IPAG wich allows 
to identify lines in the collected ASAI spectral survey using the Jet Propulsor Laboratory  
(JPL\footnote{https://spec.jpl.nasa.gov/}, Pickett et al. 1998) and Cologne Database
 for Molecular Spectroscopy (CDMS\footnote{http://www.astro.uni-koeln.de/cdms/}; 
 M\"uller et al. 2001, 2005) molecular databases. We double checked the line identifications 
 with the GILDAS Weeds package (Maret et al. 2011).
We detected several lines of H$_2^{13}$CO, HDCO, D$_2$CO, $^{13}$CH$_3$OH and CH$_2$DOH 
(see Tables \ref{table:deut} and \ref{table:13}).
Examples of the detected line profiles  in T$_{MB}$ scale are shown in Figure \ref{fig:spectra}.
The peak velocities of the detected lines are between
 +8 km s$^{-1}$ and +9 km s$^{-1}$, being consistent, once considered the fit uncertainties, with 
 the systemic velocity of both A and B component of \S (+8.6 km s$^{-1}$, Chen et al. 2009; L\'opez-Sepulcre et al. 2015).
We fitted the lines with a Gaussian function,
and excluded from the analysis those lines with
$|$v$_{peak}$ - v$_{sys}$$|>$0.6 km/s plausibly affected
by line blending.
We select for the analysis only the lines with a signal to noise (S/N) higher than 4$\sigma$.
The spectral parameters of the detected lines, as well as
the results from the Gaussian fits, are presented in 
Tables \ref{table:deut} and \ref{table:13},
where we report the frequency of each transition (GHz), 
the telescope HPBW ($\arcsec$), 
the excitation energies of the upper level $E_{\rm up}$ (K), 
 the S$\mu^{2}$ product (D$^{2}$), the line rms (mK),
  the peak temperature (mK), the peak velocities (km s$^{-1}$), the line full width at half maximum (FWHM) (km s$^{-1}$) 
  and the velocity integrated line intensity $I_{\rm int}$ (mK km s$^{-1}$).

\subsection{Formaldehyde isotopologues}\label{Results_form}

We report the detection of several lines of H$_2$CO and
 its isotopologues H$_2^{13}$CO, HDCO and D$_{2}$CO. 
The measured intensity ratio between the low energy transitions of  
 H$_2$CO and H$_2^{13}$CO (as e.g. the 3$_{\rm 1,3}$--2$_{\rm 1,2}$ at $E_{\rm up}$ = 32 K),
 is $\sim$ 25, a value well below the median value
 for the interstellar medium of $^{12}$C/$^{13}$C $\sim$ 68 (Milam et al. 2005). 
This indicates that the observed  H$_2$CO transitions are optically thick.
Therefore we use H$_2^{13}$CO to derive the formaldehyde deuteration.

We detected 7 lines of H$_2^{13}$CO, 5 lines of HDCO and 5 lines of D$_2$CO, 
with excitation energies, $E_{\rm up}$, in the 10--45 K range.
Examples of the detected line profiles are shown in Figure \ref{fig:spectra};
 the detected transitions and the observational parameters are displayed in
  Table \ref{table:deut} and Table \ref{table:13}. 
The lines profiles are close to a gaussian shape and the peak velocities are close
 to the systemic source velocity with values between 
+7.8 km s$^{-1}$ and +9.0 km s$^{-1}$ while the FWHM is between
 0.9 and 2.5 km s$^{-1}$.
Three lines of H$_2^{13}$CO (with frequencies 137.45 GHz, 141.98 GHz, and 146.64 GHz)
 and one line of HDCO (with frequency 134.2848) are detected in the 2 mm band and
 they are affected by contamination of
emission in the off positions (see Sect. 2 for details on
the observing techniques), consistently with the
 analysis reported by  L\'opez-Sepulcre et al. (2015), using ASAI spectra.
The contaminated lines correspond to a size of the telescope HPBW $>$ 16$\arcsec$.
In these cases the measured intensities will be 
treated as lower limits in the rotational diagram analysis (see Sect. \ref{sec:RD}).
For the D$_2$CO, only one line is detected at 2 mm but it does not show any
 absorption feature due to the wobbler contamination. 
This can be an indication of a more compact region emitting in D$_2$CO with respect to
 that of HDCO emission.
A similar behaviour has been observed in a different context by 
Fuente et al. (2015) towards the intermediate-mass Class 0
protostar NGC 7129ÐFIRS 2. They detected, using interferometric observations,
 an intense and compact D$_{2}$CO
component associated with the hot core.
On the other hand Ceccarelli et al. (2001) detected in the low-mass
Class 0 protostar IRAS16293-2422, an extended D$_{2}$CO emission
(up to $\sim$5000 AU), associated with the external envelope.
The present data do not allow us to draw reliable conclusions on the 
relative size of the two deuterated formaldehyde isotopologues.
However, in the case of SVS13-A a more compact size is suggested by the broader line profiles of 
D$_2$CO with respect to HDCO (see Figure \ref{fig:histo}). In Figure \ref{fig:histo}
we show the distribution of the linewidths of the detected HDCO lines in hatched blue 
and D$_2$CO lines in cyano. The bulk of the HDCO lines has a FWHM between 1.5 and 2.0 
km s$^{-1}$ while for the D$_2$CO the peak of the distribution is in the 2.0--2.5 km s$^{-1}$ range. 
A further discussion on this will be done in Section \ref{sec:Discussion} following the results of the rotational diagram analysis.

Interestingly, three lines of low excitation ($E_{\rm up}<$ 35 K) of H$_2^{13}$CO 
(with frequencies 212.81 GHz, 206.13 GHz and 219.91 GHz) and all the HDCO lines
 (except for the line in the 2 mm band) show weak ($\sim$ 30 mK) wings 
 clearly indicating emission due to outflows that we analyse separately 
 from the main line component.

\subsection{Methanol isotopologues}

Similarly to formaldehyde, the detected lines of CH$_3$OH are optically thick. 
We verified it through the measured ratio between the insensities of CH$_3$OH and 
$^{13}$CH$_3$OH (as e.g. the 5$_{\rm 1,5}$--4$_{\rm 1,4}++$ at $E_{\rm up}$ = 49 K)
 that is $\sim$ 2. For this reason also in this case, we use $^{13}$CH$_{3}$OH to 
 calculate methanol column density.
In the case of methanol, the process of line identification was more complex than formaldheyde. 
This is due to the very rich spectra observed with ASAI towards SVS13-A with a consequent
challenging lines identification for a complex molecule such as CH$_{3}$OH.
In addiction to the criteria summarized in Section \ref{sec:line_id}, we further require
FWHM $>$ 2 km s$^{-1}$ to discard any possible false identification. 
For transitions with multiple components 
(e.g. $^{13}$CH$_3$OH 2$_{\rm 1,1}$--1$_{\rm 1,0}$ and 2$_{\rm 1,1}$--1$_{\rm 1,0}$--)
we select only the lines for which the different component intensities are close to the expected LTE (Local Thermodinamic Equilibrium) relative intensities.

\begin{figure*}
\begin{center}
\includegraphics[width=17cm]{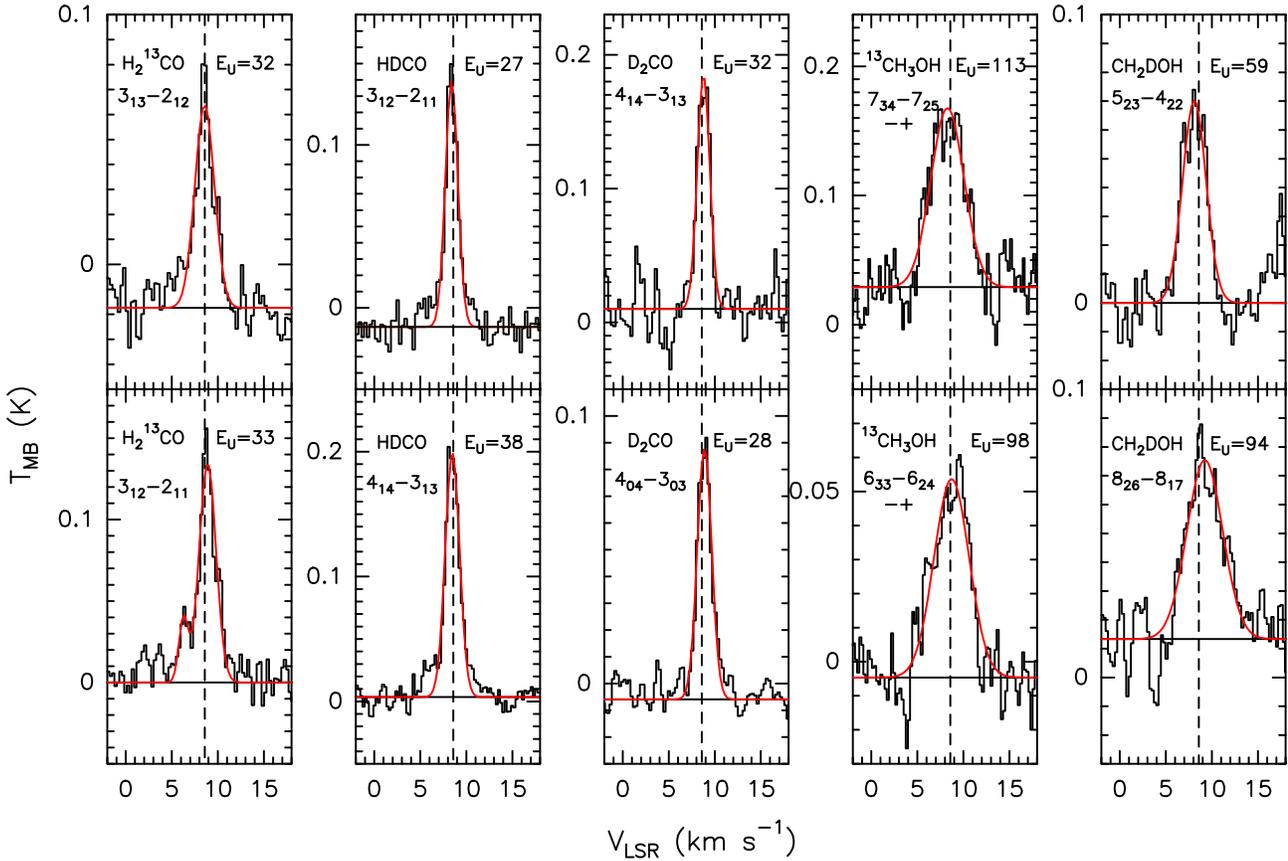}
\end{center}
\caption{Examples of line profiles  in T$_{\rm MB}$ scale (not corrected for the beam dilution): species and transitions are reported. The vertical dashed line stands for the ambient LSR velocity (+ 8.6 Km s$^{-1}$, Chen et al. 2009)} 
\label{fig:spectra}
\end{figure*}

\begin{figure}
\begin{center}
\includegraphics[angle=90,width=9cm]{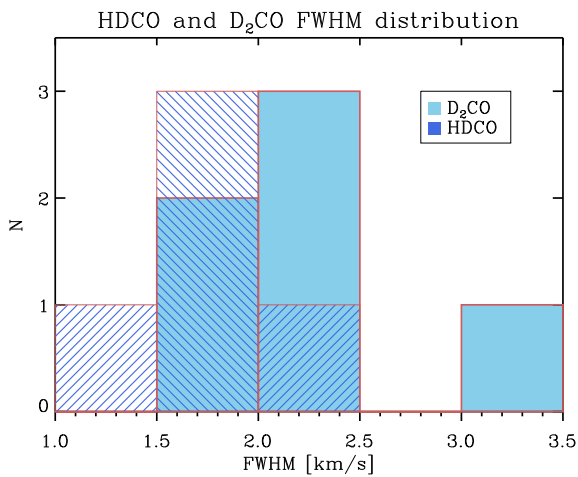}
\end{center}
\caption{Distribution of the linewidth (FWHM) of the observed HDCO and D$_2$CO lines. Cyano is for D$_2$CO and blue hatched is for HDCO.} 
\label{fig:histo}
\end{figure}

We report the detection of 18 transitions of $^{13}$CH$_3$OH and  27 lines of CH$_2$DOH with excitation energies in the 20--276 K range. 
Examples of the detected line profiles for methanol isotopologues are shown in Figure \ref{fig:spectra}.
The spectral parameters and the results of the gaussian fit are shown in Tables \ref{table:deut} and \ref{table:13}.
The line profiles are broader than for formaldehyde isotopologues, with a FWHM up to 5.4 km s$^{-1}$. 
None of the observed profiles show absorption features due to the wobbler contamination, pointing to
an emitting region smaller than formaldehyde.

\begin{figure}
\begin{center}
\includegraphics[width=6cm]{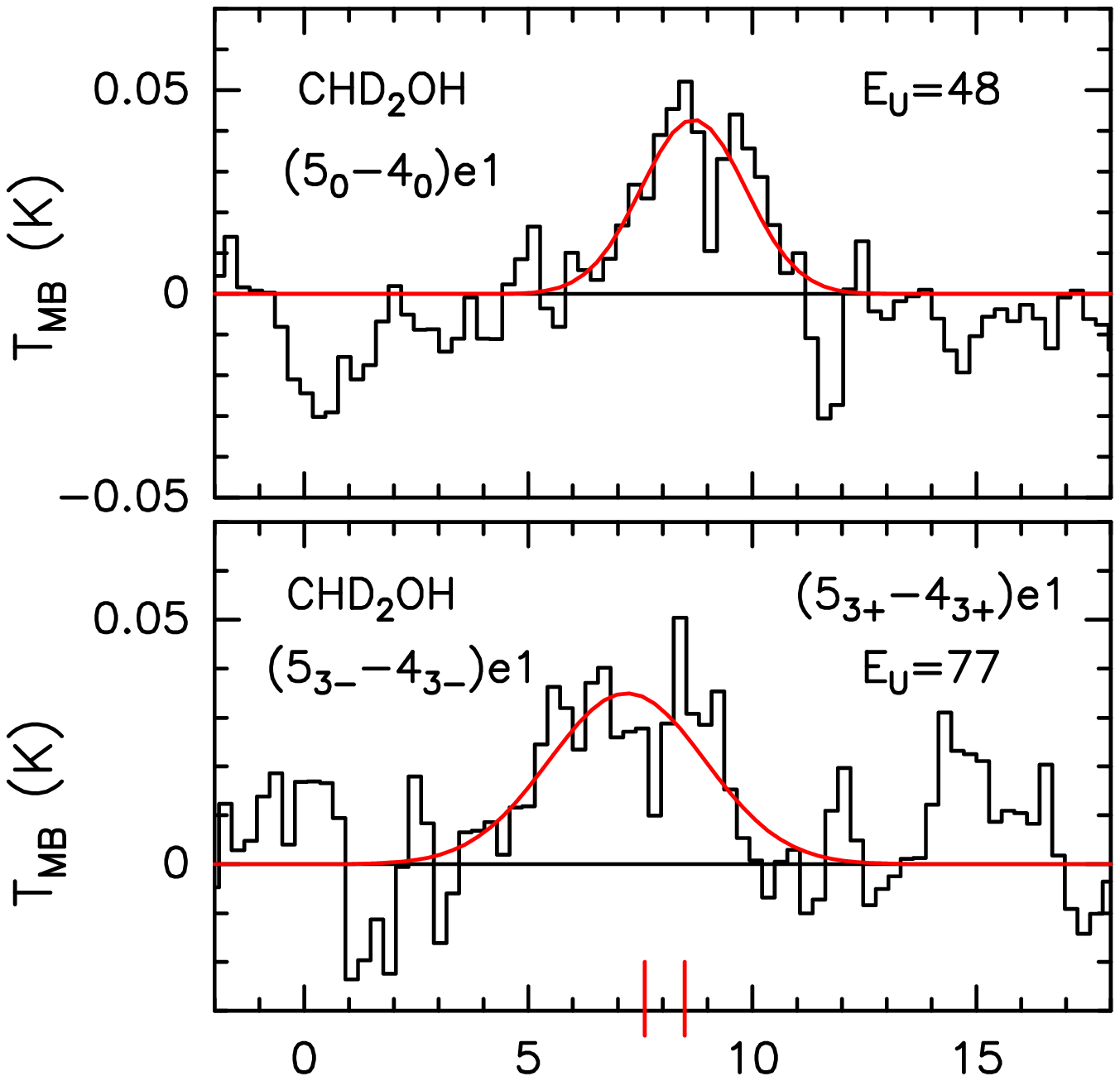}
\includegraphics[width=6cm]{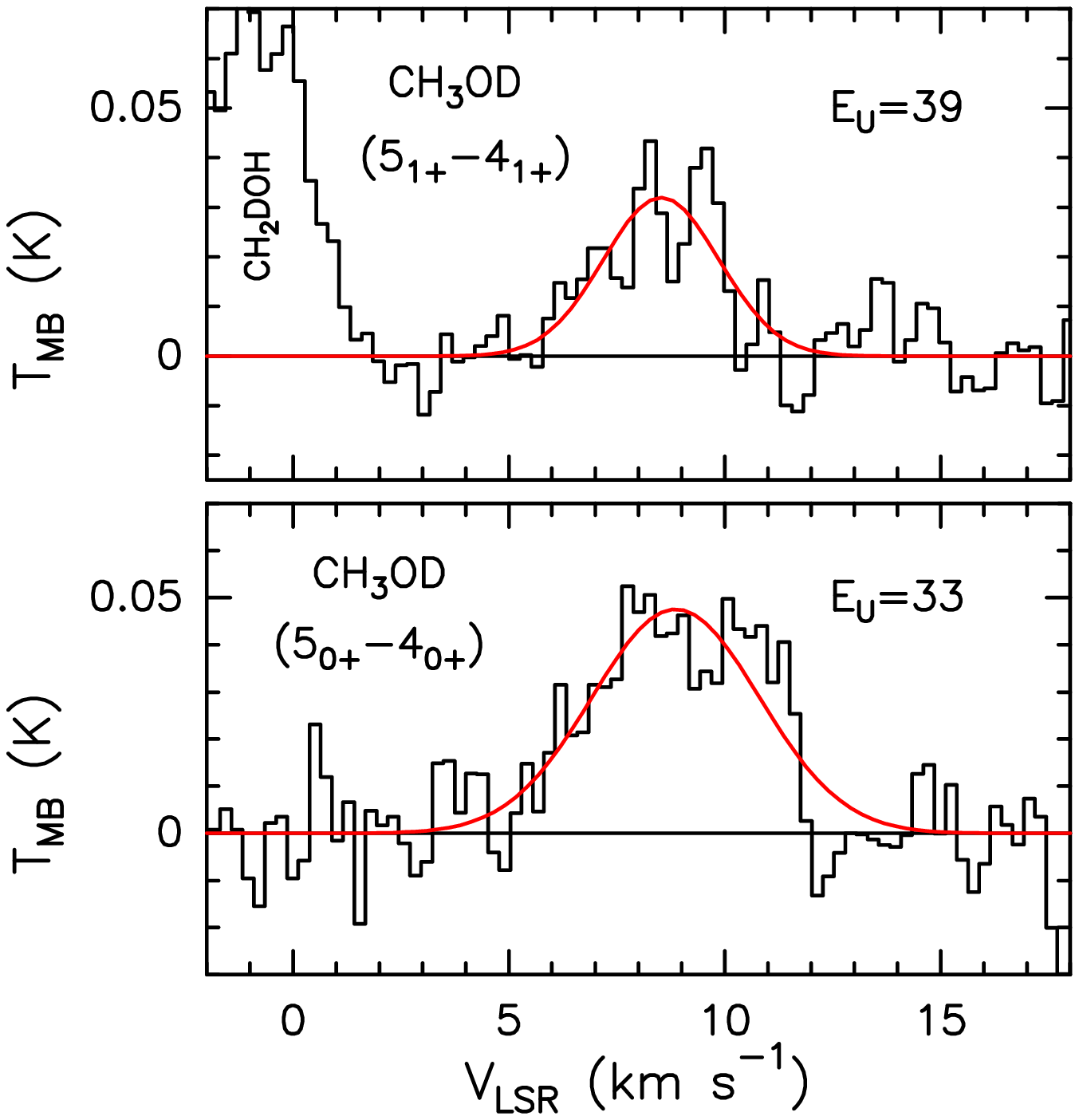}
\end{center}
\caption{Tentative detections of emission due to CHD$_2$OH (upper panel) and CH$_3$OD (lower panel) transitions.
Transitions and upper level energies are reported. Red curves are for the Gaussian fit. Note that the middle-upper panel
reports emission due to two different transitions (see the red vertical bars).}
\label{fig:tentative}
\end{figure}

Interestingly, we detect two 
different transitions of both CHD$_2$OH and CH$_3$OD with 
$E_{\rm up}$ between 33 K and 77 K (see Table 1 and Figure \ref{fig:tentative}). 
The peak velocities are consistent with the systemic source velocity and the FWHMs 
are in agreement with those of the lines from the other methanol isotopologues.

\begin{figure}
\begin{center}
\includegraphics[angle=0,width=9cm,trim=0cm(left) 0cm(bottom) 0cm(right) 0cm(up)]{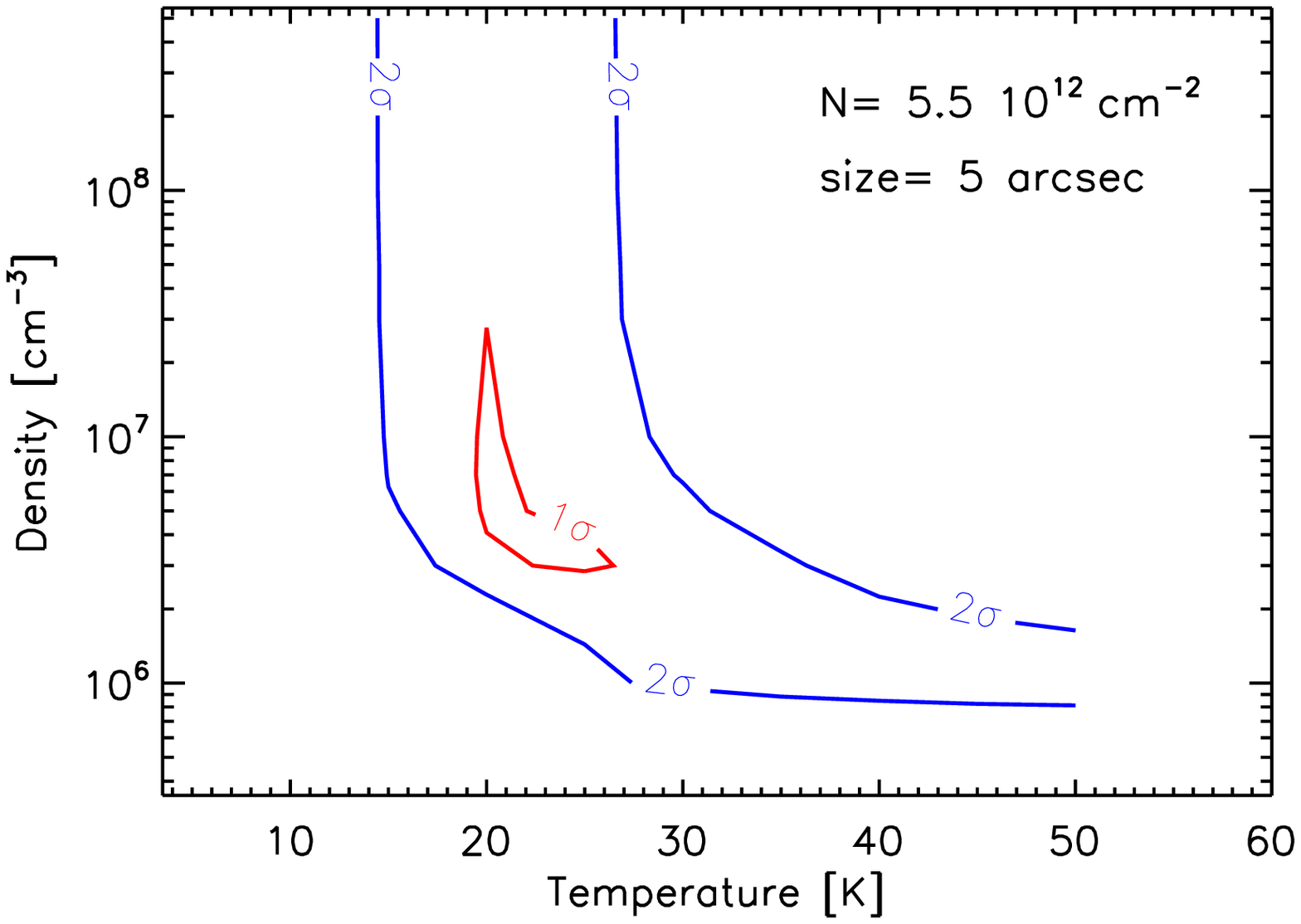}
\includegraphics[angle=0,width=9cm]{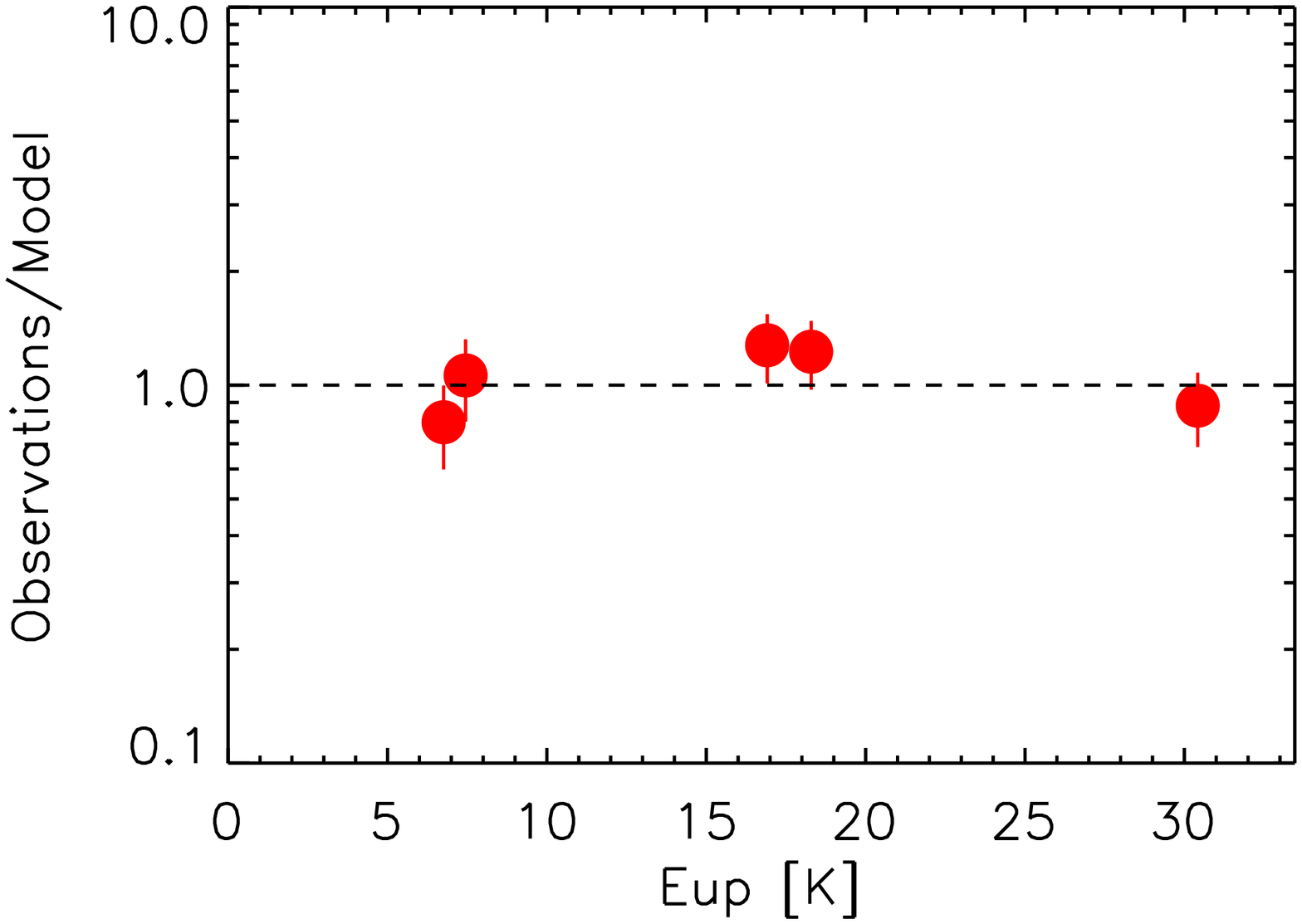}
\end{center}
\caption{ {\it Upper panel:} The 1$\sigma$ and 2$\sigma$ contour plot of  $\chi^{2}$ obtained 
considering the non-LTE model predicted and observed intensities of all
the detcted ortho $^{13}$H$_{2}$CO lines. The best fit is obtained with $N(^{13}H_{2}CO)$ = 5.5 $\times$ 10$^{12}$ cm$^{-2}$,
$\theta_{s}$ = 5 $\arcsec$, $T_{\rm kin}$ = 20 K and
$n_{\rm H_{2}}$ $\geq$ 7 $\times$ 10$^{6}$ cm$^{-3}$.
{\it Lower panel:} Ratio between the observed line intensities
with those predicted by the best fit model as a function of line
upper level energy $E_{\rm up}$.} 
\label{Fig:LVG-form}
\end{figure}

\begin{figure}
\begin{center}
\includegraphics[angle=0,width=9cm,trim=0cm(left) 0cm(bottom) 0cm(right) 0cm(up)]{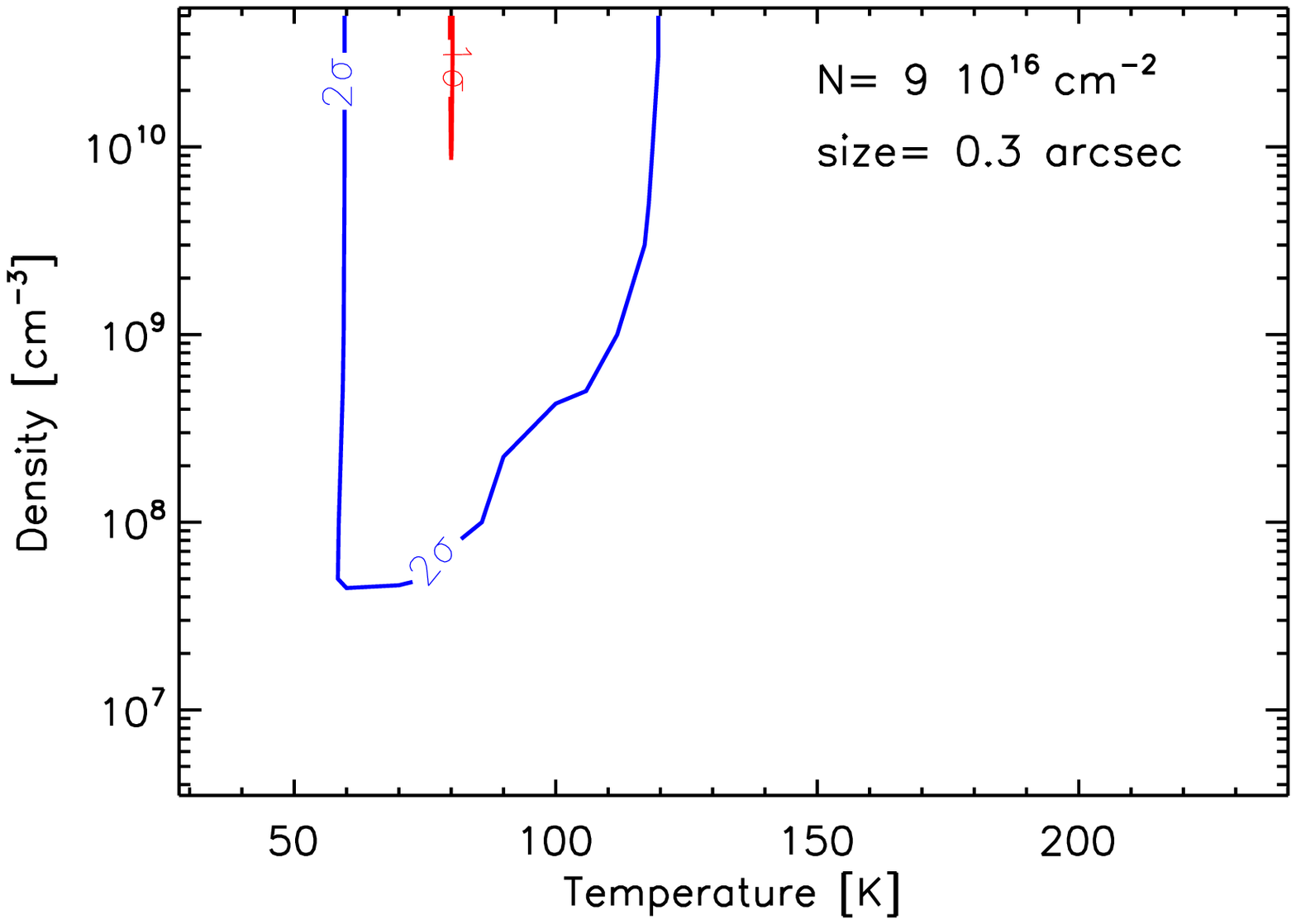}
\includegraphics[angle=0,width=9cm]{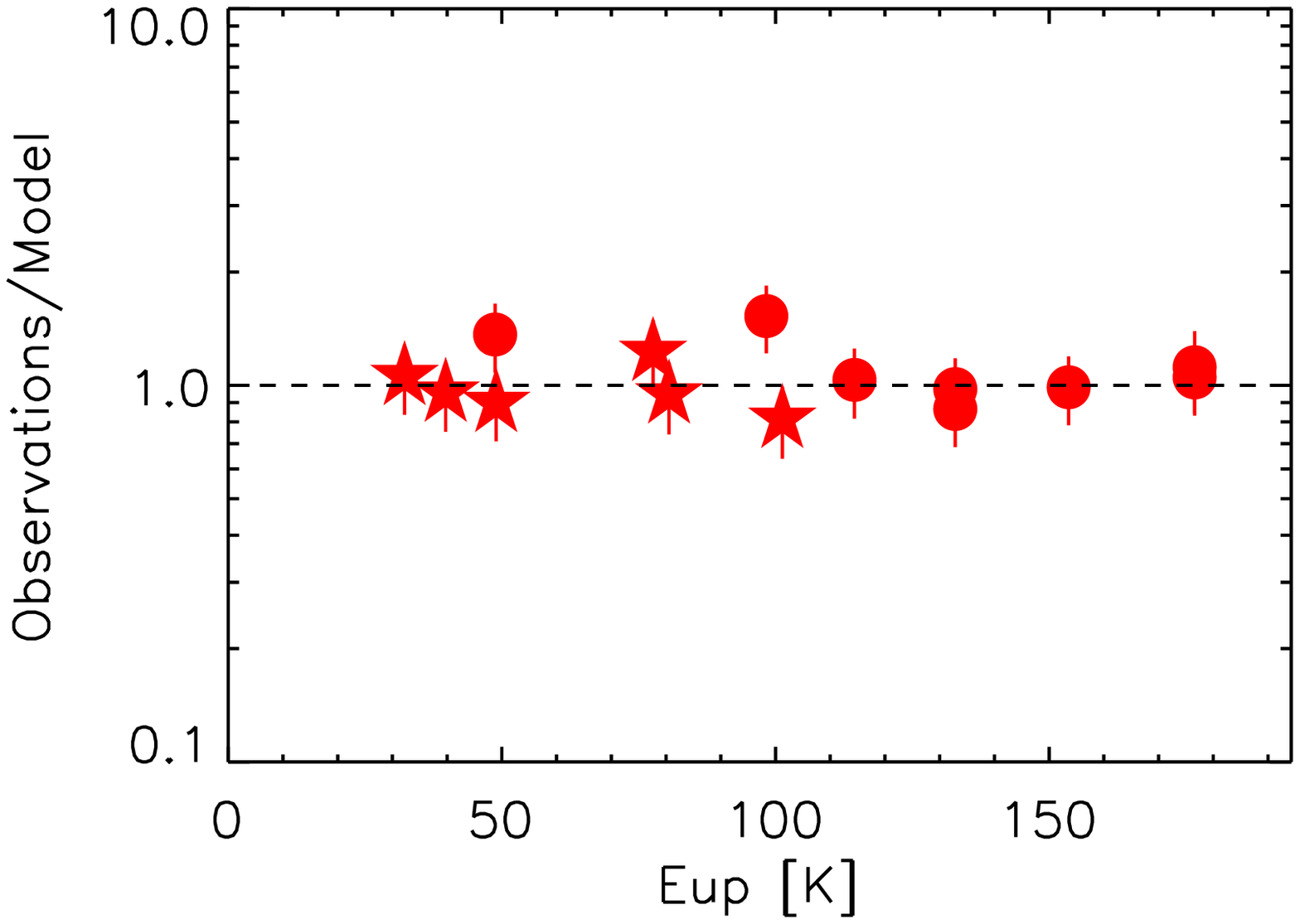}
\end{center}
\caption{ {\it Upper panel:} The 1$\sigma$ (in red) and 2$\sigma$ (in blue) contour plot of  $\chi^{2}$ obtained 
considering the non-LTE model predicted and observed intensities
the detcted $^{13}$CH$_{3}$CO lines with $E_{\rm up} >$ 40 K. The best fit is obtained with $N(^{13}CH_{3}OH)$ = 9 $\times$ 10$^{16}$ cm$^{-2}$, $\theta_{s}$ = 0$\farcs$3, $T_{\rm kin}$ = 80 K and 
$n_{\rm H_{2}}$ $\geq$ 3 $\times$ 10$^{10}$ cm$^{-3}$.
{\it Lower panel:} Ratio between the observed line intensities
with those predicted by the best fit model as a function of line
upper level energy $E_{\rm up}$. Circles refer to $^{13}$CH$_{3}$CO A transitions while stars refer to E transitions.} 
\label{Fig:LVG-met1}
\end{figure}

\begin{figure}
\begin{center}
\includegraphics[width=9cm]{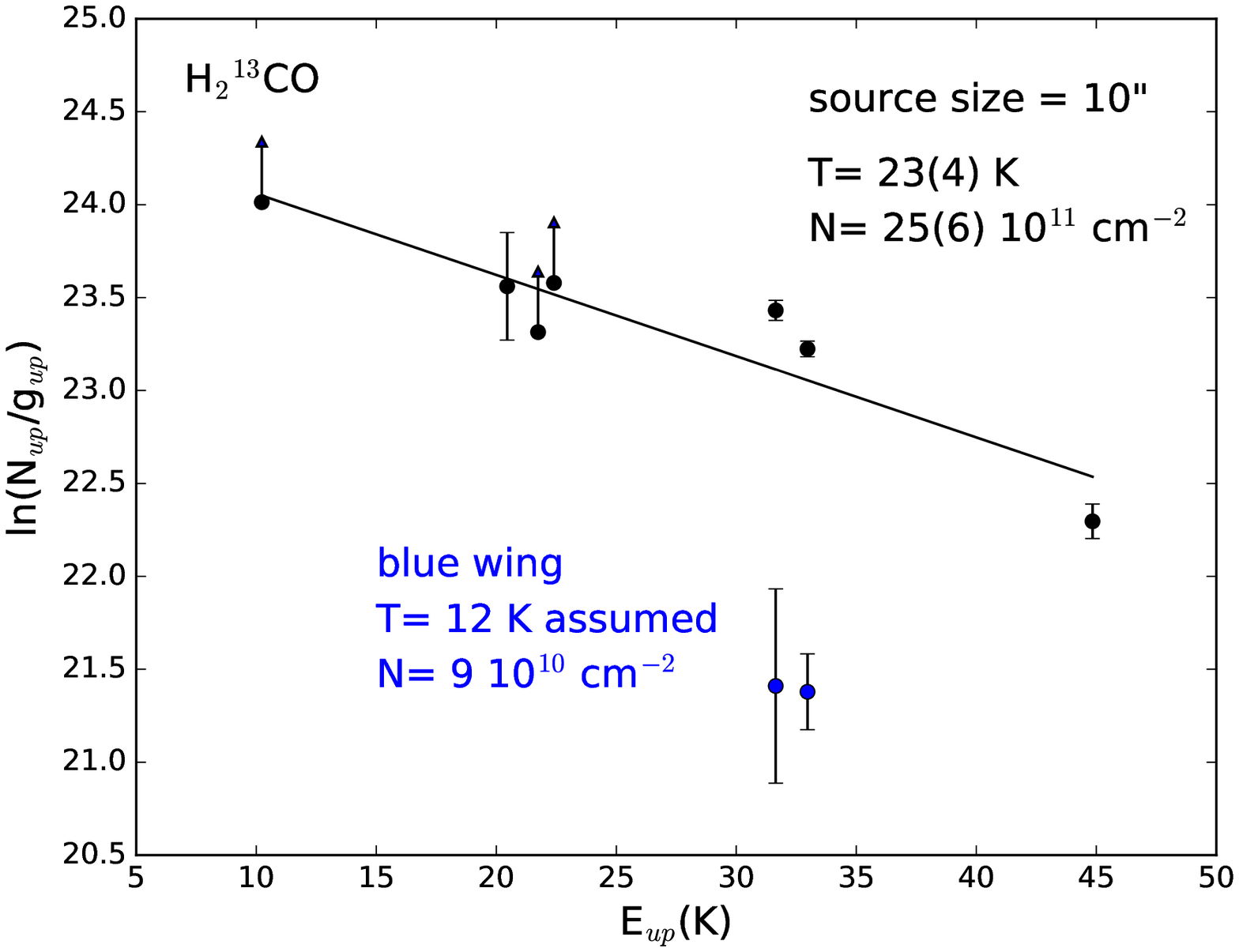}
\includegraphics[width=9cm]{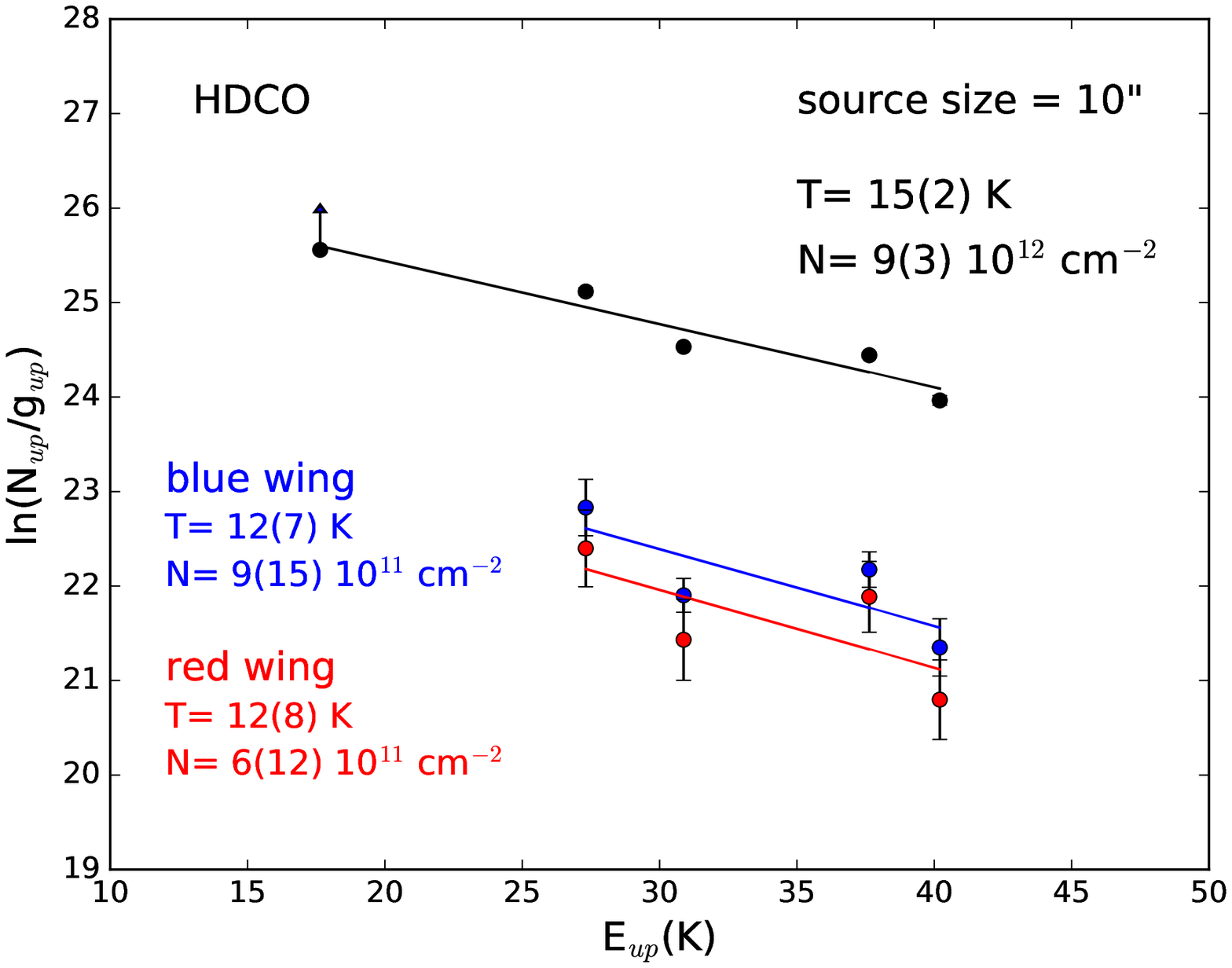}
\includegraphics[width=9cm]{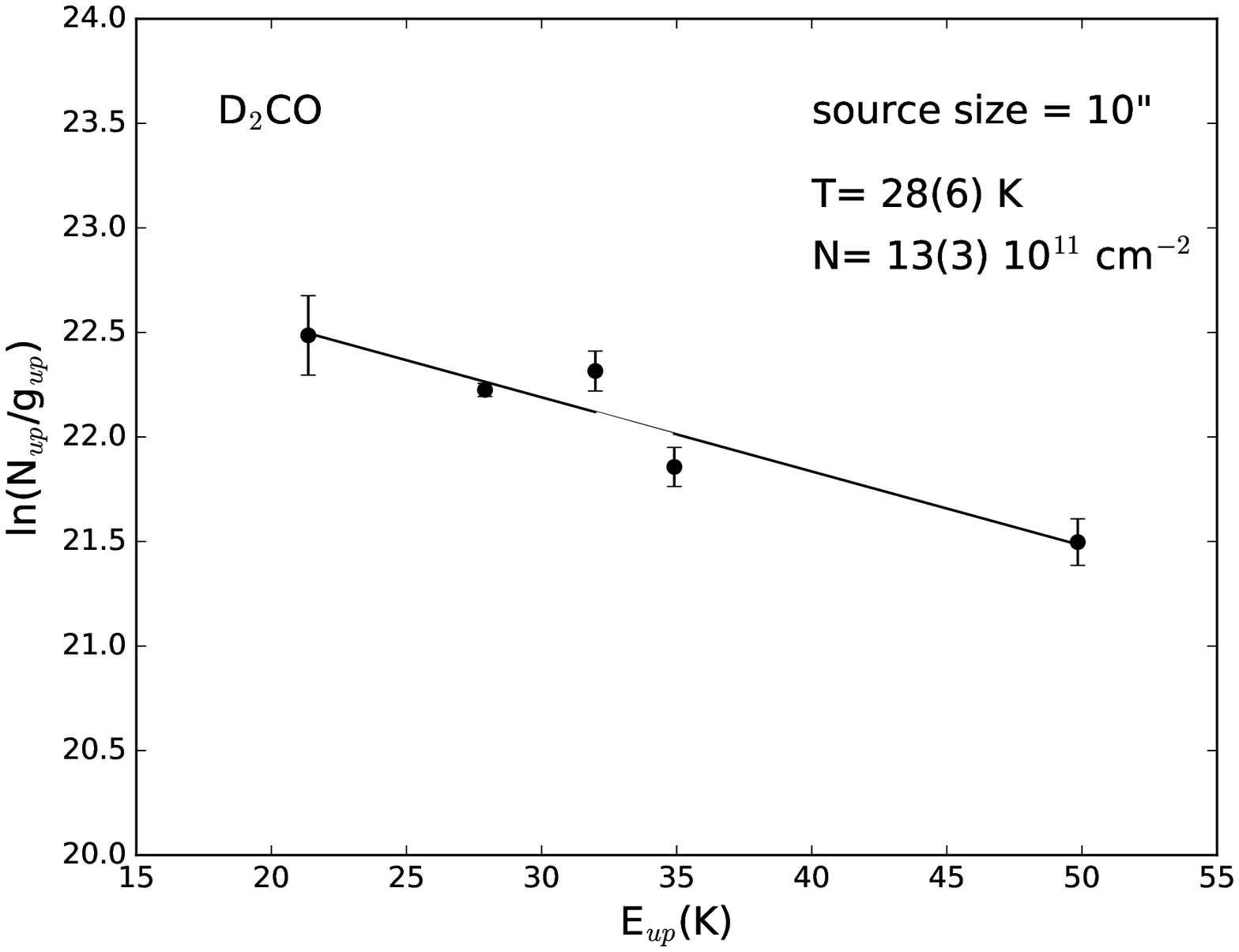}
\end{center}
\caption{Rotation diagrams for H$_2^{13}$CO (upper panel), HDCO (middle panel) and D$_2$CO (lower panel). An emitting region size of 10$\arcsec$ is assumed (see text). The parameters $N_{\rm u}$, $g_{\rm u}$, and $E_{\rm up}$ are, respectively, the column density, the degeneracy and the energy (with respect to the ground state of each symmetry) of the upper level. The derived values of the rotational temperature are reported. Arrows are for the lines affected by wobbler contamination (see Section \ref{Results_form}) and thus considered as lower limits.} 
\label{fig:RD_form}
\end{figure}

\begin{figure}
\begin{center}
\includegraphics[angle=0,width=9cm,trim=0cm(left) 0cm(bottom) 0cm(right) 0cm(up)]{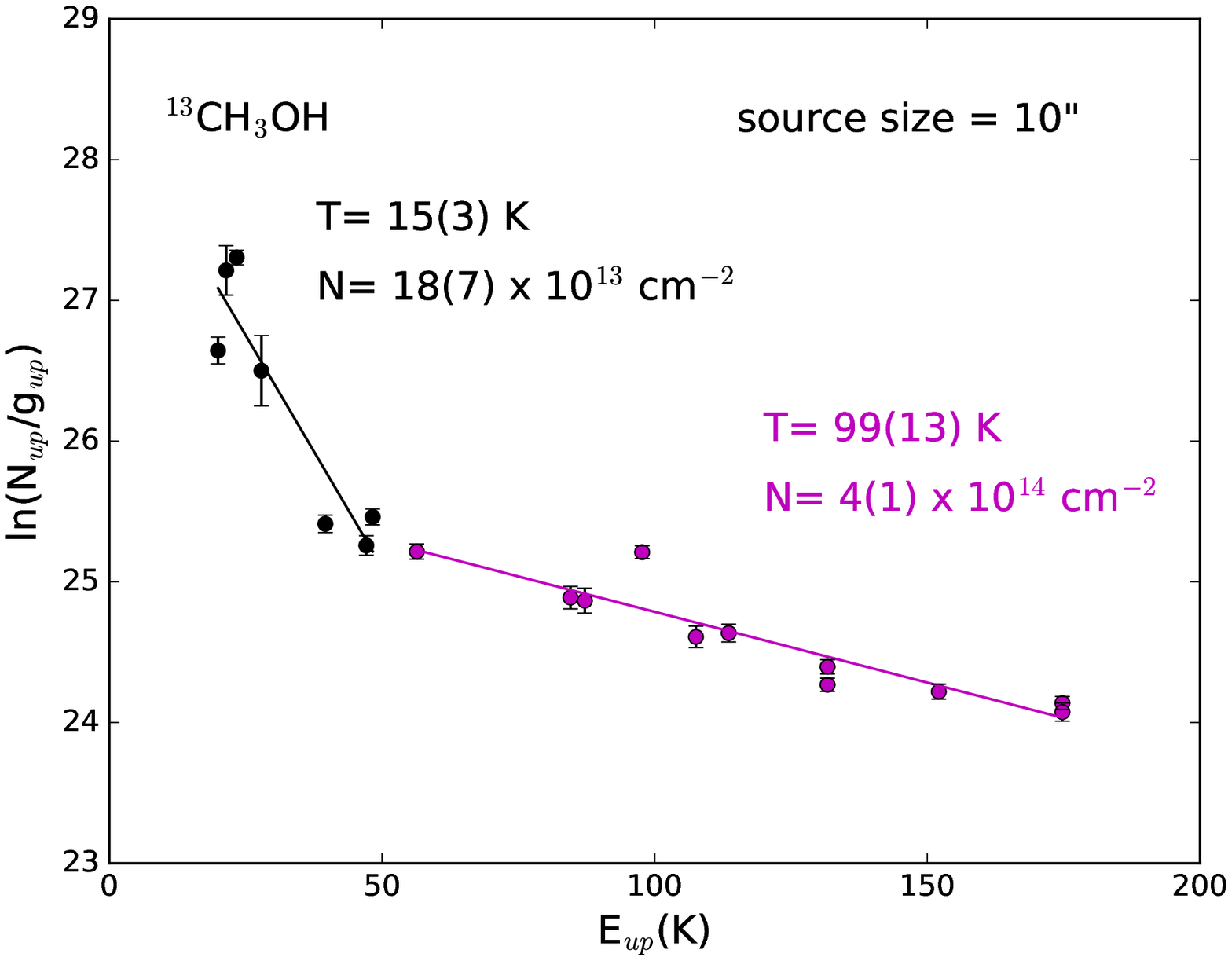}
\includegraphics[angle=0,width=9cm]{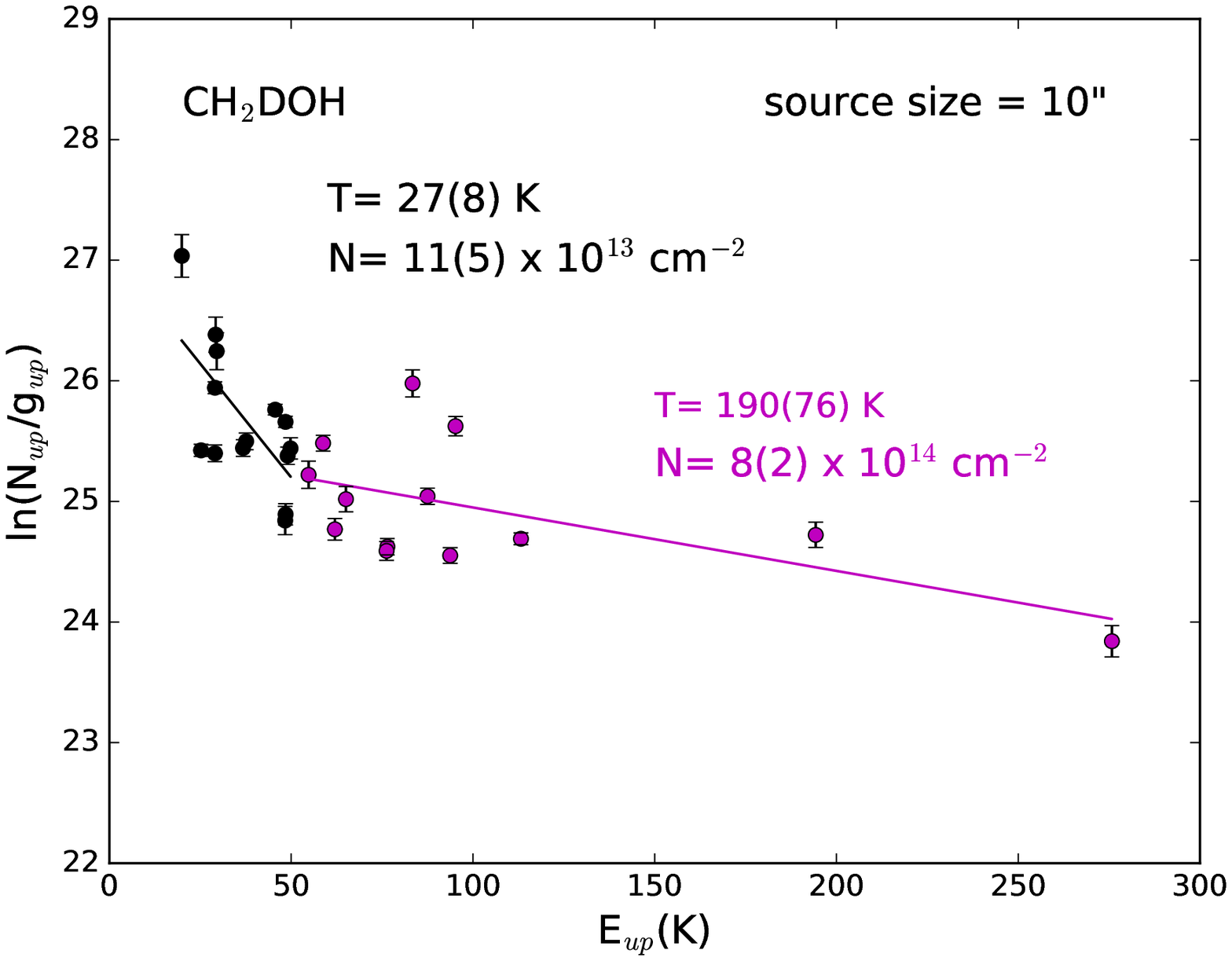}
\end{center}
\caption{Rotation diagrams for $^{13}$CH$_3$OH (upper panel) and CH$_2$DOH (lower panel) assuming two emitting components. An emitting region size of 10$\arcsec$ is assumed (see text). The parameters $N_{\rm u}$, $g_{\rm u}$, and $E_{\rm up}$ are, respectively, the column density, the degeneracy and the energy (with respect to the ground state of each symmetry) of the upper level. The derived values of the rotational temperature are reported.} 
\label{fig:RD-Met_spezzati}
\end{figure}


\begin{table*}
\caption{List of transitions and line properties (in $T_{\rm MB}$ scale) of the HDCO, D$_{\rm 2}$CO and CH$_{\rm 2}$DOH emission detected towards \SA}
\begin{tabular}{lccccccccc}

\hline
\multicolumn{1}{c}{Transition} &
\multicolumn{1}{c}{$\nu$$^{\rm a}$} &
\multicolumn{1}{c}{$HPBW$} &
\multicolumn{1}{c}{$E_{\rm up}$$^a$} &
\multicolumn{1}{c}{$S\mu^2$$^a$} &
\multicolumn{1}{c}{rms} &
\multicolumn{1}{c}{$T_{\rm peak}$$^b$} &
\multicolumn{1}{c}{$V_{\rm peak}$$^b$} &
\multicolumn{1}{c}{$FWHM$$^b$} &
\multicolumn{1}{c}{$I_{\rm int}$$^b$} \\
\multicolumn{1}{c}{ } &
\multicolumn{1}{c}{(GHz)} &
\multicolumn{1}{c}{($\arcsec$)} &
\multicolumn{1}{c}{(K)} &
\multicolumn{1}{c}{(D$^2$)} & 
\multicolumn{1}{c}{(mK)} &
\multicolumn{1}{c}{(mK)} &
\multicolumn{1}{c}{(km s$^{-1}$)} &
\multicolumn{1}{c}{(km s$^{-1}$)} &
\multicolumn{1}{c}{(mK km s$^{-1}$)} \\ 

\hline
\multicolumn{10}{c}{Deuterated species}\\
\hline

HDCO 2$_{\rm 1,1}$--1$_{\rm 1,0}$ & 134.2848 & 18 & 18 & 8   &  17 & 158(5)   & +8.31(0.05) & 1.1(0.1) & 188(17)\\
HDCO 3$_{\rm 1,2}$--2$_{\rm 1,1}$ & 201.3414 &  12 & 27 & 14  & 19 & 334(22) & +8.43(0.03) & 1.6(0.1) & 561(21)\\
HDCO 4$_{\rm 1,4}$--3$_{\rm 1,3}$ & 246.9246 &  10 & 38 & 20  & 17 & 312(22) & +8.50(0.03) & 1.9(0.1) & 619(18)\\
HDCO 4$_{\rm 0,4}$--3$_{\rm 0,3}$ & 256.5854 &  10 & 31 & 22  & 10 & 376(20) & +8.54(0.01) & 1.9(0.0) & 777(11)\\
HDCO 4$_{\rm 1,3}$--3$_{\rm 1,2}$ & 268.2920 & 9   &  40 & 20  & 21 & 207(21) & +8.55(0.05) & 2.0(0.1) & 451(23)\\

\hline

p-D$_{\rm 2}$CO 3$_{\rm 1,3}$--2$_{\rm 1,2}$ & 166.1028 & 15    &    21 &  14      & 11 & 33(9)           & +8.79(0.21) & 2.2(0.5)    & 79(15)\\
p-D$_{\rm 2}$CO 4$_{\rm 1,4}$--3$_{\rm 1,3}$ & 221.1918 & 11     &    32 &  20      & 16 & 92(7)           & +8.74(0.09) & 1.8(0.2)    & 178(17)\\
o-D$_{\rm 2}$CO 4$_{\rm 0,4}$--3$_{\rm 0,3}$ & 231.4103 & 11      &    28 &  43      & 11 & 194(12)       & +8.88(0.03) & 1.9(0.1)    & 381(12)\\


o-D$_{\rm 2}$CO 4$_{\rm 2,2}$--3$_{\rm 2,1}$ & 236.1024 & 10      &    50 &  33       & 13 & 56(7)           & +8.95(0.13) & 2.4(0.3)    & 144(16)\\
p-D$_{\rm 2}$CO 4$_{\rm 1,3}$--3$_{\rm 1,2}$ & 245.5329 & 10        &    35 &   20      & 11 & 55(7)           & +8.85(0.11) & 2.4(0.3)    & 139(13)\\

\hline


CH$_{\rm 2}$DOH 2$_{\rm 0,2}$--1$_{\rm 0,1}$ e1 & 	89.2753 & 28  & 20            & 1  &  3 & 12(2) & +8.65(0.37) & 4.2(1.1) & 51(9)\\ 

CH$_{\rm 2}$DOH 6$_{\rm 1,5}$--6$_{\rm 0,6}$ e0 & 99.6721 & 25 & 50            & 7  & 2 & 15(2) & +8.09(0.17) & 4.1(0.5) & 68(6)\\

CH$_{\rm 2}$DOH 7$_{\rm 1,6}$--7$_{\rm 0,7}$ e0 & 105.0370 & 23  &65                & 8  & 3 & 17(3) & +8.66(0.18)& 3.1(0.4) & 57(6)\\

CH$_{\rm 2}$DOH 3$_{\rm 1,2}$--2$_{\rm 1,1}$ e1 & 135.4529 & 18 & 29 &           2  & 8 & 30(6) & +8.48(0.22) & 3.0(0.5) & 98(15) \\

CH$_{\rm 2}$DOH 3$_{\rm 1,3}$--4$_{\rm 0,4}$ e1 & 161.6025 & 15 & 29 &             1 & 9 & 36(7) & +8.71(0.19) & 2.7(0.4) & 103(15)\\

CH$_{\rm 2}$DOH 5$_{\rm 1,5}$--4$_{\rm 1,4}$ o1 & 221.2730  & 11  & 55 & 4  & 17 & 57(7) & +8.31(0.19) & 3.2(0.4) & 195(22)\\ 

CH$_{\rm 2}$DOH 5$_{\rm 0,5}$--4$_{\rm 0,4}$ e1 & 222.7415 & 11  & 46 & 4 & 10 & 75(8) & +8.45(0.10) & 4.3(0.2) & 342(15)\\

\vspace{0.1 cm}

CH$_{\rm 2}$DOH 5$_{\rm 2,3}$--4$_{\rm 1,4}$ e0 & 	223.0711 & 11  & 48 & 3  & 8 & 59(9) & +7.98(0.11) & 4.5(0.2) & 284(13)\\

CH$_{\rm 2}$DOH 5$_{\rm 3,3}$--4$_{\rm 3,2}$ o1$^c$ & 223.1535 &\multirow{2}{*}{11}  &\multirow{2}{*}{87} &\multirow{2}{*}{2}&\multirow{2}{*}{10}&\multirow{2}{*}{56(7)}&\multirow{2}{*}{+8.34(0.13)}&\multirow{2}{*}{3.8(0.3)}&\multirow{2}{*}{223(15)}\\
\vspace{0.1 cm}
CH$_{\rm 2}$DOH 5$_{\rm 3,2}$--4$_{\rm 3,1}$ o1$^c$  &223.1536\\

\vspace{0.1 cm}
CH$_{\rm 2}$DOH 5$_{\rm 2,3}$--4$_{\rm 2,2}$ e1 & 223.3155 & 11 & 59 & 3  & 12 &  70(8) & +8.12(0.10) & 3.0(0.2) & 228(15)\\

CH$_{\rm 2}$DOH 5$_{\rm 4,2}$--4$_{\rm 1,1}$ e0 $^c$  & 	223.6162 & \multirow{2}{*}{11}  &\multirow{2}{*}{95} & \multirow{2}{*}{1}& \multirow{2}{*}{10} & \multirow{2}{*}{47(11)} & \multirow{2}{*}{+8.25(0.14)}& \multirow{2}{*}{3.5(0.3)} & \multirow{2}{*}{174(14)}\\
\vspace{0.1 cm}
CH$_{\rm 2}$DOH 5$_{\rm 4,1}$--4$_{\rm 0,0}$ e0 $^c$& 223.6162\\

CH$_{\rm 2}$DOH 5$_{\rm 1,4}$--4$_{\rm 1,3}$ e1 & 225.6677 & 11  & 49 & 4 & 11 & 56(11) & +8.18(0.15) & 4.0(0.3) & 237(17)\\

CH$_{\rm 2}$DOH 5$_{\rm 1,4}$--4$_{\rm 1,3}$ e0 & 	226.8183 & 11  & 37 & 3  & 10 & 51(10) & +8.08(0.14) & 3.8(0.3) & 203(14)\\

CH$_{\rm 2}$DOH 15$_{\rm 2,13}$--15$_{\rm 1,14}$ e0 & 228.2461 & 11 & 276 & 20  & 29 & 89(12) & +7.84(0.20)& 3.0(0.4) & 285(37)\\

CH$_{\rm 2}$DOH 9$_{\rm 2,7}$--9$_{\rm 1,8}$ e0 & 231.9692 & 11  & 113 & 11  & 14 & 99(8) & +8.72(0.09) & 3.7(0.2) & 390(19)\\

CH$_{\rm 2}$DOH 8$_{\rm 2,6}$--8$_{\rm 1,7}$ e0 & 234.4710 & 10 & 94 & 10  & 13 & 67(8) & +9.17(0.14) & 4.1(0.3) & 293(19)\\

CH$_{\rm 2}$DOH 7$_{\rm 2,5}$--7$_{\rm 1,6}$ e0 & 237.2499 & 10  & 76 & 8  & 12 & 64(12) & +8.31(0.13) & 3.9(0.3) & 266(18)\\

CH$_{\rm 2}$DOH 7$_{\rm 1,6}$--6$_{\rm 2,4}$ o1 & 244.5884 & 10 & 83 & 2  & 16 & 61(10) & +8.09(0.18) & 3.6(0.6) & 231(26)\\

CH$_{\rm 2}$DOH 3$_{\rm 2,1}$--3$_{\rm 1,2}$ e0 & 247.6258 & 10  & 29 & 2  & 9 & 48(8) & +8.26(0.13) & 3.7(0.3) & 189(13)\\

CH$_{\rm 2}$DOH 3$_{\rm 2,2}$--3$_{\rm 1,3}$ e0 & 255.6478 & 10 & 29 & 2 & 9 & 58(8) & +8.61(0.12) & 5.3(0.4) & 331(16)\\

CH$_{\rm 2}$DOH 4$_{\rm 1,4}$--3$_{\rm 0,3}$ e0 & 256.7316 & 10  & 25 & 3  & 9 & 61(8) & +8.32(0.11) & 4.3(0.2) & 278(14)\\

CH$_{\rm 2}$DOH 4$_{\rm 2,3}$--4$_{\rm 1,4}$ e0 & 258.3371 & 10  & 38 & 3  & 14 & 60(6) & +8.33(0.16) & 4.7(0.4) & 302(21)\\

CH$_{\rm 2}$DOH 5$_{\rm 2,4}$--5$_{\rm 1,5}$ e0 & 	261.6874 & 9  & 48 & 4 & 17 & 45(9) & +8.09(0.26) & 4.3(0.5) & 205(24)\\

CH$_{\rm 2}$DOH 13$_{\rm 0,13}$--12$_{\rm 1,12}$ e0 & 262.5969 & 9  & 194 & 5 & 17 & 54(12) & +8.51(0.21) & 3.8(0.4) & 219(23)\\

CH$_{\rm 2}$DOH 6$_{\rm 1,6}$--5$_{\rm 1,5}$ e0 & 264.0177 & 9  & 48 & 4 & 14 & 64(7) & +8.40(0.12) & 2.8(0.3) & 192(17)\\

CH$_{\rm 2}$DOH 7$_{\rm 2,6}$--7$_{\rm 1,7}$ e0 & 	270.2999 & 9  & 76 & 6  & 14 & 55(13) & +8.35(0.17) & 4.2(0.4) & 243(19)\\

CH$_{\rm 2}$DOH 6$_{\rm 1,5}$--5$_{\rm 1,4}$ e1 & 270.7346 & 9  & 62 & 4 	 & 16 & 60(10) & +8.21(0.16) & 3.5(0.3) & 222(20)\\

\hline
\vspace{0.1 cm}
CHD$_{\rm 2}$OH 5$_{\rm 0}$--4$_{\rm 0}$ e1 & 207.771 & 11  & 48 & 4 	 & 14 & 43(9) & +8.69(0.20) & 2.7(0.4) & 125(17)\\

CHD$_{\rm 2}$OH 5$_{\rm 3-}$--4$_{\rm 3-}$ e1$^c$  & 207.868 &\multirow{2}{*}{11}  &\multirow{2}{*}{77} &\multirow{2}{*}{2}&\multirow{2}{*}{11}&\multirow{2}{*}{35(10)}&\multirow{2}{*}{+7.21(0.24)}&\multirow{2}{*}{4.1(0.5)}&\multirow{2}{*}{153(17)}\\
\vspace{0.1 cm}
CHD$_{\rm 2}$OH 5$_{\rm 3-}$--4$_{\rm 3-}$ e1$^c$  & 207.869 \\

\hline

CH$_{\rm 3}$OD 5$_{\rm 1+}$--4$_{\rm 1+}$ & 223.3086 & 11  & 39 & 3 	 & 20 & 32(10) & +8.53(0.41) & 3.2(0.9) & 108(27)\\

CH$_{\rm 3}$OD 5$_{\rm 0+}$--4$_{\rm 0+}$  & 226.5387 & 11  & 33 & 4 	 & 18 & 48(10) & +8.87(0.28) & 4.6(0.6) & 232(28)\\

\hline

\end{tabular}

$^a$ Frequencies and spectroscopic parameters of HDCO and D$_{\rm 2}$CO have been extracted from the Cologne Database for Molecular Spectroscopy (M\"uller et al. 2005). Those of CH$_{\rm 2}$DOH are extracted from the Jet Propulsion Laboratory database (Pickett et al. 1998).
$^b$ The errors in brackets are the gaussian fit uncertainties.
$^c$ The lines cannot be distinguished with the present spectral resolution. \\
\label{table:deut}
\end{table*}



\begin{table*}

  \caption{List of transitions and line properties (in $T_{\rm MB}$ scale) of the H$_{\rm 2}^{\rm 13}$CO and $^{\rm 13}$CH$_{\rm 3}$OH emission towards \SA}
\begin{tabular}{lccccccccccc}
\hline
\multicolumn{1}{c}{Transition} &
\multicolumn{1}{c}{$\nu$$^{\rm a}$} &
\multicolumn{1}{c}{$HPBW$} &
\multicolumn{1}{c}{$E_{\rm up}$$^a$} &
\multicolumn{1}{c}{$S\mu^2$$^a$} &
\multicolumn{1}{c}{rms} &
\multicolumn{1}{c}{$T_{\rm peak}$$^b$} &
\multicolumn{1}{c}{$V_{\rm peak}$$^b$} &
\multicolumn{1}{c}{$FWHM$$^b$} &
\multicolumn{1}{c}{$I_{\rm int}$$^b$} \\
\multicolumn{1}{c}{ } &
\multicolumn{1}{c}{(GHz)} &
\multicolumn{1}{c}{($\arcsec$)} &
\multicolumn{1}{c}{(K)} &
\multicolumn{1}{c}{(D$^2$)} & 
\multicolumn{1}{c}{(mK)} &
\multicolumn{1}{c}{(mK)} &
\multicolumn{1}{c}{(km s$^{-1}$)} &
\multicolumn{1}{c}{(km s$^{-1}$)} &
\multicolumn{1}{c}{(mK km s$^{-1}$)} \\

\hline
\multicolumn{10}{c}{Isotopologues}\\
\hline


o-H$_{\rm 2}^{\rm 13}$CO 2$_{\rm 1,2}$--1$_{\rm 1,1}$ & 137.4500 & 18  & 22 & 24  & 10 & 63(4) & +8.50(0.08) & 1.0(0.2) & 64 (10)\\
p-H$_{\rm 2}^{\rm 13}$CO 2$_{\rm 0,2}$--1$_{\rm 0,1}$ & 141.9837  & 17 & 10 & 11  & 10 & 54 (10) & +8.46(0.10)  & 1.1(0.2)& 62(10) \\
o-H$_{\rm 2}^{\rm 13}$CO 2$_{\rm 1,1}$--1$_{\rm 1,0}$ & 146.6357 & 17  & 22 & 24 & 14 & 105(4) & +8.39(0.06)  & 0.9(0.1) & 98(13)\\
o-H$_{\rm 2}^{\rm 13}$CO 3$_{\rm 1,3}$--2$_{\rm 1,2}$ & 206.1316 & 12  & 32 & 44  & 13 & 124(15) & +8.58(0.06) & 2.5(0.2) & 332(18) \\
p-H$_{\rm 2}^{\rm 13}$CO 3$_{\rm 0,3}$--2$_{\rm 0,2}$ & 212.8112 & 12 & 20 & 16 & 10 & 70(6) & +7.76(0.14) & 2.0(0.3) & 235(65) \\
o-H$_{\rm 2}^{\rm 13}$CO 3$_{\rm 1,2}$--2$_{\rm 1,1}$ & 219.9085 & 11  &33 & 43  & 10 & 134(12) & +8.90(0.04) & 2.2(0.1) & 359(17) \\
o-H$_{\rm 2}^{\rm 13}$CO 4$_{\rm 1,4}$--3$_{\rm 1,3}$ & 274.7621 & 10  & 45 & 31  & 21 & 102(16) & +8.34(0.11) & 2.5(0.3) & 270(25) \\

\hline

$^{\rm 13}$CH$_{\rm 3}$OH 2$_{\rm 0,2}$--1$_{\rm 0,1}$ & 94.4110 & 26  &  20 & 2  & 2 & 10(2)  & +8.61(0.21) & 3.9 (0.4) &  42(4) \\
$^{\rm 13}$CH$_{\rm 3}$OH 2$_{\rm 1,1}$--1$_{\rm 1,0}$ & 94.4205  & 26  &28 & 1  & 2 & 11(1) & +9.19(0.16) & 2.4(0.4)& 27(4) \\
$^{\rm 13}$CH$_{\rm 3}$OH 2$_{\rm 1,1}$--1$_{\rm 1,0}$-- & 95.2087 & 26  & 21 & 1  & 2 & 17(2) & +7.56(0.10) & 3.5(0.3) & 65(4) \\

$^{\rm 13}$CH$_{\rm 3}$OH 1$_{\rm 1,0}$--1$_{\rm 0,1}$ & 165.5661 & 15  & 23 & 1  & 8 & 76(8) & +8.58(0.09) & 3.6(0.2)& 289(15) \\
$^{\rm 13}$CH$_{\rm 3}$OH 7$_{\rm 1,6}$--7$_{\rm 0,7}$ & 166.5695 & 15 & 84 & 6 & 5 & 34(5) & +8.64(0.13) & 3.5(0.4) & 125(10) \\

$^{\rm 13}$CH$_{\rm 3}$OH 8$_{\rm -1,8}$--7$_{\rm 0,7}$ & 221.2852  & 11& 87 & 5  & 11 & 49(11) & +8.05(0.16) & 3.5(0.3) & 180(16) \\
$^{\rm 13}$CH$_{\rm 3}$OH 5$_{\rm 1,5}$--4$_{\rm 1,4}$++ & 234.0116 & 11  & 48 & 4  & 10 & 66(13) & +8.87(0.11) & 4.1(0.3) & 284(16) \\
$^{\rm 13}$CH$_{\rm 3}$OH 5$_{\rm 0,5}$--4$_{\rm 0,4}$ & 235.8812 & 10  & 47 & 4  & 13 & 70(8) & +9.11(0.12) & 3.3(0.3)& 245(17) \\
$^{\rm 13}$CH$_{\rm 3}$OH 5$_{\rm -1,5}$--4$_{\rm -1,4}$ & 235.9382 & 10  & 40 & 4 & 10 & 52(10) &+8.84(0.15) & 5.0(0.3)& 275(17) \\
$^{\rm 13}$CH$_{\rm 3}$OH 10$_{\rm 3,7}$--10$_{\rm 2,8}$-+ & 254.5094 & 10  & 175 & 9  & 8 & 54(7) & +8.31(0.09) & 3.7(0.2) & 215(10) \\
$^{\rm 13}$CH$_{\rm 3}$OH 8$_{\rm 3,5}$--8$_{\rm 2,6}$-+ & 254.8418 & 10  & 132 & 7  & 7 & 52(7) &+8.29(0.09) &4.0(0.2) & 218(11) \\
$^{\rm 13}$CH$_{\rm 3}$OH 7$_{\rm 3,4}$--7$_{\rm 2,5}$-+ & 254.9594 & 10 & 113 & 6  & 11 & 52(7) & +8.25(0.14)& 4.3(0.3) & 238(15) \\
$^{\rm 13}$CH$_{\rm 3}$OH 6$_{\rm 3,3}$--6$_{\rm 2,4}$-+ & 255.0510 & 10  & 98 & 5 & 10 & 71(11) & +8.63(0.11) & 4.7(0.2)& 353(16) \\
$^{\rm 13}$CH$_{\rm 3}$OH 8$_{\rm 3,6}$--8$_{\rm 2,7}$+- & 255.2656 & 10  &132 & 7  & 6 & 47(6) & +8.30(0.09) & 3.9(0.2)& 193(9) \\
$^{\rm 13}$CH$_{\rm 3}$OH 9$_{\rm 3,7}$--9$_{\rm 2,8}$+- & 255.3559 & 10  &152 & 8  & 7 & 51(7) & +8.52(0.09) & 3.8(0.2)& 208(11) \\
$^{\rm 13}$CH$_{\rm 3}$OH 10$_{\rm 3,8}$--10$_{\rm 2,9}$+- & 255.4970 & 10  & 175 & 9  & 9 & 46(9) & +8.43(0.14)  & 4.1(0.3) & 202(13)\\
$^{\rm 13}$CH$_{\rm 3}$OH 5$_{\rm 2,3}$--4$_{\rm 1,3}$ & 263.1133 & 9 & 56 &  4  & 10 & 60(10) & +8.31(0.12) & 4.4(0.3)& 278(15) \\
$^{\rm 13}$CH$_{\rm 3}$OH 9$_{\rm -1,9}$--8$_{\rm 0,8}$ & 268.6354 & 9  &107 & 6  & 13 & 55(13) & +7.99(0.16)  & 4.1(0.4)& 236(18)\\

\hline
\end{tabular}

$^a$ Frequencies and spectroscopic parameters of H$_{\rm 2}^{\rm 13}$CO and $^{\rm 13}$CH$_{\rm 3}$OH have been extracted from the Cologne Database for Molecular Spectroscopy (M\"uller et al. 2005). Upper level energies refer to the corresponding ground state of each symmetry.
$^b$ The errors in brackets are the gaussian fit uncertainties. \\
\label{table:13}
\end{table*}

\subsection{Summary of the results}

In summary, the bulk of methanol and formaldehyde isotopologues lines
are detected in the 1 mm band. For this reason, the temperature estimate
 from the rotational diagram analysis (see Section \ref{sec:RD})
is not affected by the beam dilution.
The 30-m HPBW is $\sim$10$\arcsec$ at 1 mm, which ensures that the emission is coming from SVS13-A with
no contamination from SVS13-B (the separation between SVS13-A and the companion protostar is $\sim$ 15$\arcsec$).
The lines colllected in the 2 and 3 mm bands could be contaminated by the emission from SVS13-B,
because the HPBW is larger, but they are only a handful of lines.

 Interestingly, the formaldehyde profiles show line wings that suggest emission
due to the extended outflow driven by SVS13-A ($>$ 0.07 pc, Lefloch et al. 1998, Codella et
al. 1999).

\section{Discussion}\label{sec:Discussion}

\subsection{LVG analysis}

We analysed the H$_{2}$$^{13}$CO and $^{13}$CH$_{3}$OH observed lines with the non-
LTE Large Velocity Gradient (LVG) approach using the model described in Ceccarelli et al. (2003). 
For methanol we used the CH$_{3}$OH-H$_{2}$ collisional coefficients
provided by the BASECOL database (Dubernet et al. 2013).
In the case of formaldehyde, we considered only the ortho form, 
for which the H$_{2}$CO-H$_{2}$ collisional coefficients (Troscompt et al. 2009a) are available. 
We assumed a Boltzmann distribution for the H$_{2}$, using for the methanol
analysis the statistical ortho-to-para ratio of 3. In the case of formaldehyde 
we assumed a ortho-to-para ratio close to zero following Troscompt et al. (2009b).
We ran grids of models varying the kinetic temperature, $T_{\rm kin}$, (from
10 to 200 K),  
the H$_{2}$ density, $n_{\rm H_{2}}$, (from 10$^{4}$ to 10$^{10}$ cm$^{-3}$), 
the H$_{2}$$^{13}$CO column density, $N(^{13}H_{2}CO)$, (from
10$^{11}$ to 10$^{13}$ cm$^{-2}$), and the $^{13}$CH$_3$OH 
column density, $N(^{13}CH_{3}OH)$, (from 10$^{16}$ to 10$^{18}$ cm$^{-2}$), while
the emitting size, $\theta_{s}$, was left as free parameter.

In the case of formaldehyde, the best fit was obtained with 
$N(^{13}H_{2}CO)$ = 5.5 $\times$ 10$^{12}$ cm$^{-2}$ and 
$\theta_{s}$ = 5$\arcsec \pm$ 1$\arcsec$: Figure \ref{Fig:LVG-form} (upper panel) 
shows the $\chi^{2}_{r}$ contour plot as a function of the
temperature and H$_{2}$ density using these values. The temperatures corresponding
to the best fit solution are $T_{\rm kin}$ = 20-25 K and the density are quite high  
$n_{\rm H_{2}}$ $\simeq$ 0.2--2 $\times$ 10$^{7}$ cm$^{-3}$, suggesting to be close to LTE.
Figure \ref{Fig:LVG-form} (lower panel) shows, for the best fit solution, the ratio
between the measured lines intensities and the
LVG model predictions, as a function of the line upper level energy.
The detected transitions are predicted to be optically thin (opacities§ between 0.03 and 0.06).
The LVG analysis clearly supports the association of formaldehyde with the
protostellar envelope with a size of $\sim$ 1200 AU.

Different is the case of $^{13}$CH$_{3}$OH for which the LVG model
does not converge towards a solution suggesting we are mixing emission
from different regions, possibly due to different HPBWs.
Following this suggestion, we considered separately the lines 
with higher excitations ($E_{\rm up} >$ 40 K) observed 
with similar HPBWs (between 9$\arcsec$ and 15$\arcsec$).
The solution with the lowest $\chi^{2}_{r}$ corresponds to 
$N(^{13}CH_{3}OH)$ = 9 $\times$ 10$^{16}$ cm$^{-2}$
and an emitting size of $\theta_{s}$ = 0$\farcs$3 $\pm$ 0$\farcs$1, i.e. a radius of 35 AU (see Figure \ref{Fig:LVG-met1}). 
The best fit
solution corresponds to a temperature of $T_{kin}$ = 80 K and very high-densities, 
$n_{\rm H_{2}}$ $\geq$ 10$^{8}$ cm$^{-3}$.
The line opacities vary from 0.8 to 2.5, being thus moderately optically thin.
All these values 
suggest that the emission detected at high excitations is dominated by a hot corino,
an environment which is typically very abundant in methanol, due to thermal
evaporation of the dust mantles (e.g. Caselli \& Ceccarelli 2012).
Interestingly, the occurrence of a hot-corino around SVS13-A has been recently
suggested by high-excitation HDO lines, also observed in the ASAI context
and indicating a $T_{\rm kin}$ larger than 150 K on smaller spatial scales (a radius $\sim$ 25 AU; Codella et al. 2016). 

The analysis of the remaining 4 lines (observed with HPBWs larger than 15$\arcsec$) is not straightforward given the 4 transitions have almost the same
$E_{\rm up}$ (20 -- 28 K).
The LVG approach suggests typical solutions with column densities of 
$N(^{13}CH_{3}OH)$ $\sim$ 10$^{15}$ cm$^{-2}$, temperatures $\leq$ 70 K,
densities at least 10$^{6}$ cm$^{-3}$, and sizes $\simeq$ 2$\arcsec$ -- 4$\arcsec$.
The line opacities in this case range from 0.007 to 0.02, being thus optically thin.
The lower densities and the more extended emitting size suggest that we are sampling 
a more extended region (a radius $\sim$ 350 AU) around the protostar where the temperature is still  high enough
to allow the methanol molecules to be released from grain mantles.

\subsection{Rotational Diagram analysis}\label{sec:RD}

The LVG analysis previously described suggests LTE
conditions and optically thin lines. 
As a consequence we used the 
rotational diagram analysis to 
determine the temperature 
and the column density of formaldehyde and methanol isotopologues 
through a more direct approach.
For a given molecule, the relative population distribution of 
all the energy levels, is described by a Boltzmann temperature, that is the rotational temperature $T_{\rm rot}$.
The upper level column density can be written as:

\begin{equation} 
N_{u} = \frac{8 \pi k \nu^{2}}{h c^{3} A_{ul}} \frac{1}{\eta_{bf}} \int T_{mb} dV
\label{eq:col_dens}
\end{equation}

where k is the Boltzmann constant, $\nu$ is the frequency of the transition, h is the Plank constant, c is the light speed, A$ul$ is the Einstein coefficient, $\eta_{bf}$\footnote{$\eta_{bf} = \theta_{s}^{2} \times (\theta_{s}^{2}+\theta_{b}^{2})^{-1}$; $\theta_{s}$ and $\theta_{b}$ are the source and the beam sizes (assumed to be both a circular Gaussian).} is the beam-filling factor and the integral is the integrated line intensities.

$N_{\rm u}$ is related to the rotational temperature $T_{\rm rot}$, as follow:

\begin{equation} 
ln \frac{N_{u}}{g_{u}} = ln N_{tot} - ln Q(T_{rot}) - \frac{E_{up}}{k T_{rot}} 
\end{equation}

where $g_{\rm u}$ is the generacy of the upper level, $N_{\rm tot}$ is the total column density of the molecule, Q($T_{\rm rot}$) is the partition function at the rotational temperature and $E_{\rm up}$ is the energy of the upper level.

As a first step we assumed a size filling the smaller IRAM 30-m beam,
i.e. 10$\arcsec$, a value consistent with the continuum
emission at 1.25 mm observed with IRAM 30-m radiotelescope
by Lefloch et al. (1998).
Note however that the $T_{\rm rot}$ estimate
 does not depend on the source size assumption because almost all the lines
have been observed with a beam of $\sim$ 10$\arcsec$
and then suffer the same beam dilution.
The rotational diagram analysis shows low values of $T_{\rm rot}$, around 20 K,
 consistent with the LVG results and consistent with  
an association with 
the extended molecular envelope around the protostar.
We obtained $T_{\rm rot}$ = 23 $\pm$ 4 K and column density 
 $N_{\rm tot}$ = 25 $\pm$ 6 $\times$ 10$^{11}$ cm$^{-2}$ (H$_2^{13}$CO),
 $T_{\rm rot}$ = 15 $\pm$ 2 K and $N_{\rm tot}$ = 9 $\pm$ 3 $\times$ 10$^{12}$ cm$^{-2}$ (HDCO), and
 $T_{\rm rot}$ = 28 $\pm$ 6 K and column density 
 $N_{\rm tot}$ = 13 $\pm$ 3 $\times$ 10$^{11}$ cm$^{-2}$ (D$_2$CO), see Figure \ref{fig:RD_form}.

For HDCO we detected line wings with velocities up to $\sim \pm$3 km s$^{-1}$ 
with respect to the systemic source velocity. This low velocity emission
is likely probing ambient material swept-up by the outflow associated with SVS13-A (Lefloch et al. 1998).
We derived the temperature and column density of this outflow component 
using the residual intensities after subtracting the gaussian fit of 
the ambient component and then we analysed them separately. 
From the rotational diagram analysis we obtained for both the blue- 
and the red-shifted emission, a $T_{\rm rot}$ $\sim$ 12 K.
Also in this case, the $T_{\rm rot}$ value is not affected by beam 
dilution because the lines come from the 1.3 mm band.
The low $T_{\rm rot}$ value is again an indication of an extended emission, in agreement 
with the well-studied extended outflow driven by SVS13-A (Lefloch et al. 1998).
We assumed also in this case an arbitrary source size of 10$\arcsec$,
 obtaining  $T_{\rm rot}$ = 12$\pm$ 7 K and 
 $N_{\rm tot}$ = 9$\pm$ 15 $\times$ 10$^{11}$cm$^{-2}$ (HDCO blue wing),
 $T_{\rm rot}$ = 12$\pm$ 8 K and 
 $N_{\rm tot}$ = 6$\pm$ 12 $\times$ 10$^{11}$cm$^{-2}$ (HDCO red wing).
In the case of H$_2^{13}$CO, due to line contamination, we detected blue wings
only for two lines; by assuming the same rotational temperature of the HDCO wings,
we obtained a column density of
$N_{\rm tot}$ $\sim$ 9 $\times$ 10$^{10}$cm$^{-2}$.

For H$_2^{13}$CO and HDCO we detect both para and ortho transitions (see Tables \ref{table:deut} and \ref{table:13}).
Once considered both species in a single rotation diagram, the distribution does not show any significant scatter from the linear fit.
Considering the poor statistic (2 para and 5 ortho transitions for H$_2^{13}$CO;
3 para and 2 ortho transitions for HDCO) and the uncertainties of the line intensities,
this is consistent with the o/p statistical values at the high-temperature limit (3:1 for H$_2^{13}$CO
and 2:1 for D$_2$CO).

For the methanol analysis one rotational temperature is not able to fit the rotational diagrams of $^{13}$CH$_3$OH and CH$_2$DOH, supporting the occurrence of two emitting components
associated with different excitation conditions, as already suggested by the LVG analysis.

 A better fit is obtained using two slopes (see Figure \ref{fig:RD-Met_spezzati}; again as a first step assuming 
a source size of 10$\arcsec$):
\begin{enumerate}

\item
one with a low $T_{\rm rot}$ (15 $\pm$ 3 K for $^{13}$CH$_3$OH and 27 $\pm$ 8 K for CH$_2$DOH)
for the lines with $E_{\rm up} <$ 50 K.
The column densities are $N_{\rm tot}$ = 18 $\pm$ 7 $\times$ 10$^{13}$cm$^{-2}$ for $^{13}$CH$_3$OH and $N_{\rm tot}$ = 11 $\pm$ 5 $\times$ 10$^{13}$cm$^{-2}$ for CH$_2$DOH;
\\
\item
one with a higher $T_{\rm rot}$ (99 $\pm$ 13 K for $^{13}$CH$_3$OH and 190 $\pm$ 76 K for CH$_2$DOH)
for the lines with $E_{\rm up} >$ 50 K.
The column densities are $N_{\rm tot}$ = 4 $\pm$ 1 $\times$ 10$^{14}$cm$^{-2}$ for $^{13}$CH$_3$OH and $N_{\rm tot}$ = 8 $\pm$ 2 $\times$ 10$^{14}$cm$^{-2}$ for CH$_2$DOH.

\end{enumerate}

These two excitation regimes are in agreement with what found with the LVG approach:
a hot corino and a more extended region associated with a lower temperature.
The higher $T_{\rm rot}$ values obtained for both $^{13}$CH$_3$OH 
and CH$_2$DOH with respect to the formaldehyde isotopologues suggest again that 
the origin of the emission is not the extended envelope but the hot corino.

\subsection{Methanol and formaldehyde deuteration}

We use the column densities derived from the rotation diagrams to derive the D/H ratio for formaldehyde and methanol. In order to properly measure the D/H, the column densities are
derived assuming for each species, the source size suggested by the LVG analysis:
5$\arcsec$ for formaldehyde isotopologues, $\sim$ 3$\arcsec$ for methanol lines with 
$E_{\rm up} <$ 50 K and 0$\farcs$3 for methanol lines with 
$E_{\rm up} >$ 50 K.
As already discuss in Section \ref{sec:Results}, it was not possible to directly
measure the column density of the main isotopologue of H$_2$CO and CH$_3$OH 
because the lines are optically thick. For this reason we derived the formaldehyde 
and methanol column densities from the
 H$_2^{13}$CO and $^{13}$CH$_3$OH column densities, 
 assuming a $^{12}$C/$^{13}$C ratio of 86 (Milam et al. 2005)
 at the galactocentric distance of SVS13-A.

We report the obtained D/H ratios in Table \ref{tab:RD}. 
To be consistent, we assumed for the D-species the $T_{\rm rot}$ derived from the $^{13}$C-isotopologues. In any case the following conclusions do not change if we assume for all the molecules the corresponding $T_{\rm rot}$.

For H$_2$CO we measured a D/H of 9 $\pm$ 4 $\times$ 10$^{-2}$. 
We can compare this value with measurements of deuterated formaldehyde
 in Class 0 sources performed by Parise et al. (2006), using data obtained 
 with the same antenna (IRAM 30-m) and a consistent beam sampling.
The value measured towards SVS13-A is close to the average value 
reported for the Class 0 sources, which is D/H $\sim$ 0.12. 

For the double deuterated formaldehyde we obtained a D/H value of 
4 $\pm$ 1 $\times$ 10$^{-3}$.  If we compare this value with that 
reported by Parise et al. (2006), we can note that it is definitely lower, 
by at least one order of magnitude, suggesting that the D/H is indeed lower in the more evolved 
Class I objects, like SVS13-A, with respect to the Class 0 sources.

The D/H value for the D$_2$CO with respect to the HDCO,
 is $\sim$ 5 $\times$ 10$^{-3}$, a value again lower of at least 
one order of magnitude that those reported by Parise et al. (2006) 
for the Class 0 sources. This estimate is even more reliable
 because it is independent from H$_2^{13}$CO.

Finally, we derived the D/H ratio also for the outflowing gas.
In this case, we assumed an extended component with a source size 
of 10$\arcsec$, obtaining a value of 4 $\pm$ 6 $\times$ 10$^{-3}$ 
for the HDCO in the blue wing and 3 $\pm$ 6 $\times$ 10$^{-3}$ 
for the HDCO in the red wing. These measurements are in agreement 
with that measured in the shocked region associated with
the L1157 protostellar outflow by Codella et al. (2012), 
that reported a value of 5--8 $\times$ 10$^{-3}$ using IRAM 30-m data.

\begin{figure}
\begin{center}
\includegraphics[angle=0,width=9cm,trim=0cm(left) 0cm(bottom) 0cm(right) 0cm(up)]{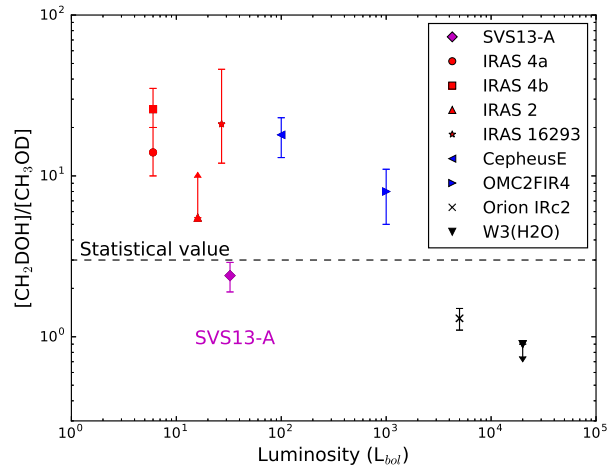}
\end{center}
\caption{Adapted from Ratajczak et al. (2011). The figure show the 
[CH$_{2}$DOH]/[CH$_{3}$OD] ratio as a function of the protostar luminosity.
 The horizontal dashed line refers to the value predicted by grain chemistry models (Charnley et al. 1997).} 
\label{CHD2OH/CH3OD}
\end{figure}

The derived D/H ratio for CH$_2$DOH with respect to CH$_3$OH is indicated 
in Table \ref{tab:RD}. To calculate this ratio, we derived the CH$_2$DOH 
column density assuming the same $T_{\rm rot}$ of $^{13}$CH$_3$OH,
obtaining D/H $\sim$ 2 $\times$ 10$^{-3}$, 
for the lines with excitation energies $E_{\rm up} <$ 50 K, 
and D/H $\sim$ 7 $\pm$ 1 $\times$ 10$^{-3}$ for the lines with $E_{\rm up}>$ 50 K.
These values are two orders of magnitude below the D/H reported in Parise et al. (2006),
supporting that also the methanol deuteration for the Class I object SVS13-A 
is dramatically decreased with respect to Class 0 objects.

We give an estimate of the CHD$_2$OH and CH$_3$OD column densities using the tentative detected two lines, 
which can be used as lower limits for the following analysis.
We derived a value of $N_{\rm tot} \sim$ 1 $\times$ 10$^{16}$ cm$^{-2}$ for CHD$_2$OH and 
$N_{\rm tot} \sim$ 6 $\times$ 10$^{14}$ cm$^{-2}$ for CH$_3$OD, assuming 
the same source size and  $T_{\rm rot}$ of the $^{13}$CH$_3$OH low energy transitions
(size $\sim$ 3$\arcsec$ and $T_{\rm rot}$ = 12 K) and using the rotational partition functions from Ratajczak et al. (2011).

\subsection{The [CH$_2$DOH]/[CH$_3$OD] ratio}

Finally, we used the detection of CH$_3$OD to derive a measure of the [CH$_2$DOH]/[CH$_3$OD]
ratio, and thus test the predictions of the current theory of methanol deuteration.
Basically, according to the grain chemistry statistical models of Charnley et al. (1997) and Osamura et al. (2004) 
the ratio of the singly deutereted isotopologues CH$_2$DOH and CH$_3$OD formed on the mantles 
should always be 3. However, this is not confirmed by the few measurements in 
star forming regions.

Figure 9 (from Ratajczak et al. 2011 and reference therein) reports the so far measured ratios as a function 
of the bolometric luminosity, including both low- and high-mass star forming regions. 
The [CH$_2$DOH]/[CH$_3$OD] ratio always differs from the statistical value suggesting a weak trend: the abundance ratio is substantially lower in massive hot cores than in (low-mass) hot-corinos (as well as in intermediate-mass protostars), by typically one order of magnitude.
In particular, in low mass protostars, CH$_3$OD is found to be less abundant than CH$_2$DOH,
by more than a factor 10 (Ratajczak et al. 2011).
Unless the prediction for the methanol formation on dust grains has to be revised, 
these measurements are suggesting that the ratio is altered 
by gas-phase reactions at work once the deuterated methanol molecules are released 
by the dust mantles.

The present work allows us to provide a little piece of information to this general context.
For SVS13-A we obtained [CH$_2$DOH]/[CH$_3$OD] in the 2.0 -- 2.5 range 
(see the magenta point in Figure 9),
comparing the column density estimated from the CH$_3$OD 5$_{\rm 1+}$--4$_{\rm 1+}$ line
and the column density from a CH$_2$DOH line with similar energy (4$_{\rm 2,3}$--4$_{\rm 1,4}$ e0).
Our measurement seems to question the previous conclusions on a change 
of the [CH$_2$DOH]/[CH$_3$OD] ratio as a function of the protostellar luminosity.
On the other hand, it suggests an evolution with time going from Class 0 to Class I, with 
CH$_2$DOH more efficiently destroyed than CH$_3$OD. To conclude, it is clear that we 
need further measurements to properly investigare any possible dependence on time and/or
luminosity.

\begin{figure*}
\begin{center}
\includegraphics[angle=0,width=17cm]{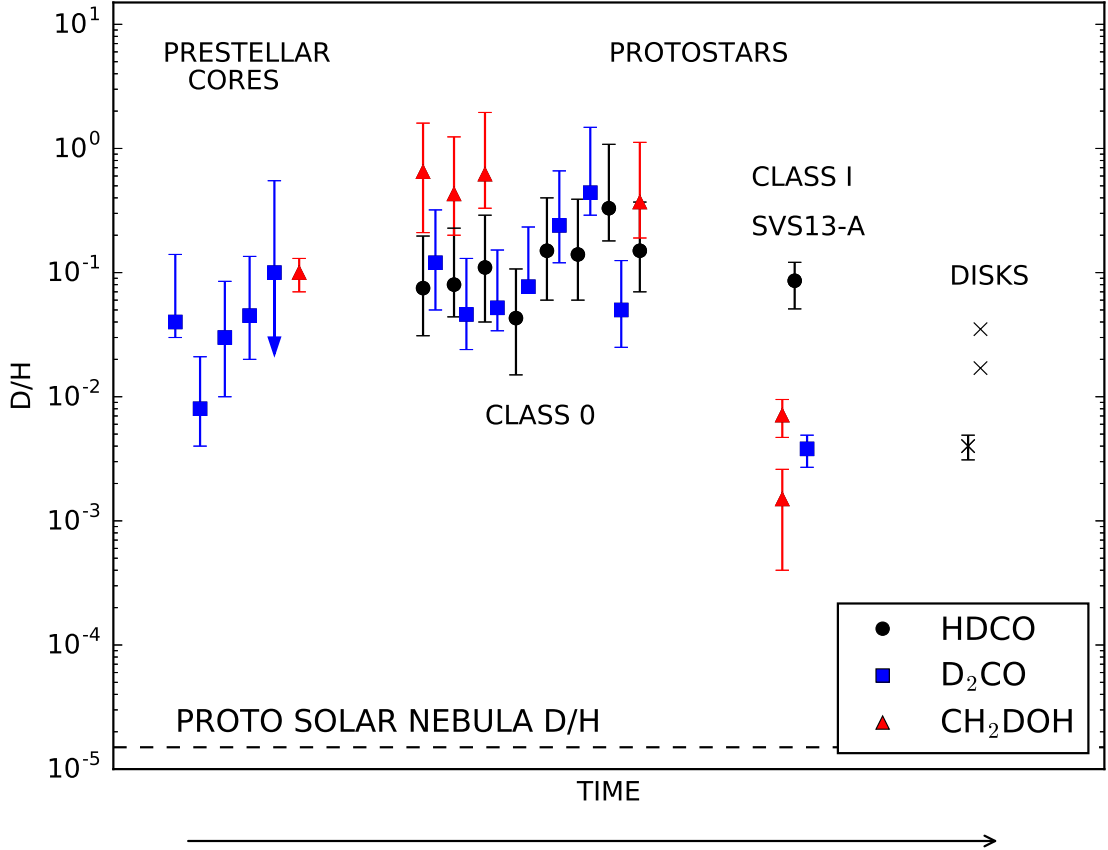}
\end{center}
\caption{D/H ratio measured in organic matter in different astronomical sources. 
Prestellar cores measurements of D$_{2}$CO and CH$_2$DOH are from respectively 
Bacmann et al. (2003) and Bizzocchi et al. (2014). 
Class 0 data are taken from Parise et al. (2006).
SVS13-A data refer to the D/H inferred in the present paper 
for HDCO (8.6 $\times$ 10$^{-2}$), D$_{2}$CO (3.8 $\times$ 10$^{-3}$) and
CH$_2$DOH (7.1 $\times$ 10$^{-3}$ for the hot corino and 1.5 $\times$ 10$^{-3}$ for a larger region, i.e. a radius $\leq$ 350 AU). 
Protoplanetary disks data refer to measurements of DCN/HCN (Oberg et al. 2012) and DCO$^{+}$/HCO$^{+}$ (van Dishoeck et al. 2006) in TW Hya and of DCO$^{+}$/HCO$^{+}$ (Guilloteau et al. 2006) in DM Tau.
Note that the prestellar cores
and the Class 0 protostars are not ordered in age thus any trend whithin the classes is not significant (as in Ceccarelli et al. 2014).}

\label{Deut}
\end{figure*}

\subsection{Deuterium fractionation of organics: from Class 0 to Class I}

The present results strongly support that both H$_{2}$CO and CH$_{3}$OH deuteration
decreases when a protostar leaves the Class 0 stage to enter
in the Class I phase.
Figure \ref{Deut} shows the D/H ratio measured for organic molecules
at different stages of the Sun-like star forming process, from
prestellar cores to protoplanetary disks
(the time increases from the left to the right along the x-axis).
The present observation for SVS13-A can be properly
compared with that of Class 0 objects,
derived by sampling
similar spatial scales around the protostar. The methanol and formaldehyde
deuteration measurements of SVS13-A, fill in the gap between
Class 0 objects and protoplanetary disks, associated with Class II-III objects.

For HDCO, the average value measured
in Class 0 sources (Parise et al. 2006) is D/H $\sim$ 0.12,
consistent with the value measured in SVS13-A, which is
D/H $\sim$ 8.6 $\pm$ 3.5 $\times$ 10$^{-2}$.
Completely different is the case of D$_{2}$CO, which  
shows an increase going from prestellar cores (average value
D/H $\sim$ 0.045, Bacmann et al. 2003) to Class 0 sources
(D/H $\sim$ 0.15, Parise et al. 2006) and then a strong decrease in
SVS13-A  (D/H = 3.8 $\pm$ 1.1 $\times$ 10$^{-3}$.
A similar behaviour is observed for the methanol deuteration that
increase from a value of D/H $\sim$ 0.1 in prestellar cores (Bizzocchi et al. 2014)
to D/H $\sim$ 0.52 in Class 0 (Parise et al. 2006) and then
significantly decrease in SVS13-A to D/H = (1.5 -- 7.1) $\times$ 10$^{-3}$.

In conclusion, the overall comparison shows a clear trend
going from the prestellar cores to the Class 0 objects and to the Class I source.
The deuterium fractionation of organics increase
going from prestellar cores to Class 0 sources and then decreases
up to two orders of magnitude going from Class 0
protostars to the more evolved phases.
In protoplanetary disks the few available organics measurements refer to
DCN/HCN (Oberg et al. 2012) and DCO$^{+}$/HCO$^{+}$ (van Dishoeck et al 2006, Guilloteau et al. 2006)
and are in agreement with the decreasing trend with values between 0.035 and 0.004.
Note that the prestellar cores
and the Class 0 protostars are not ordered in age 
thus any trend whithin the classes is not significant (as in Ceccarelli et al. 2014).


\begin{table*}
\caption{Results from the rotational diagram analysis: derived rotational temperatures, $T_{\rm rot}$, derived column densities, $N_{\rm tot}$, and  resulting deuteration ratios. The latter are calculated assuming for each deuterated species, the same $T_{\rm rot}$ of the correspondent 13-isotopologue.}
\begin{tabular}{lcccccc}
\hline
\hline
\multicolumn{1}{c}{Transition} &
\multicolumn{1}{c}{Lines} &
\multicolumn{1}{c}{Energy range} &
\multicolumn{3}{c}{Boltzmann Plots} &
\multicolumn{1}{c}{ $D/H$$^{b}$} \\
\cline{4-6}
\multicolumn{1}{c}{} &
\multicolumn{1}{c}{} &
\multicolumn{1}{c}{} &
\multicolumn{1}{c}{Size$^{a}$} &
\multicolumn{1}{c}{T$_{\rm rot}$} &
\multicolumn{1}{c}{N$_{\rm tot}$} &
\multicolumn{1}{c}{} \\
\multicolumn{1}{c}{} &
\multicolumn{1}{c}{} &
\multicolumn{1}{c}{(K) } &
\multicolumn{1}{c}{('') } &
\multicolumn{1}{c}{(K)} &
\multicolumn{1}{c}{(cm$^{-2}$)} &
\multicolumn{1}{c}{ }\\ 
\hline

\multicolumn{7}{c}{whole emission}\\ 

\hline

D$_{\rm 2}$CO & 5 & 21--50 & 5 & 25(5) & 3(1) $\times$ 10$^{12}$ & 3.8(1.1) $\times$ 10$^{-3}$\\

HDCO & 5 & 18--40 & 5 & 12(2)  &  3(1) $\times$ 10$^{13}$ & 8.6(3.5) $\times$ 10$^{-2}$\\

H$_{\rm 2}^{\rm 13}$CO & 7 & 10--45 & 5 &   19(3) & 7(2) $\times$ 10$^{12}$ & --\\

\hline

CH$_{\rm 2}$DOH ($E_{\rm up}<$ 50 K) &14 & 20--50 & $\sim$ 3 & 24(9) & 7(5) $\times$ 10$^{14}$ & 1.5(1.1) $\times$ 10$^{-3}$\\

$^{\rm 13}$CH$_{\rm 3}$OH ($E_{\rm up} <$ 50 K)& 7 & 20--48 & $\sim$ 3 & 12(2) & 16(7) $\times$ 10$^{14}$ & -- \\

CH$_{\rm 2}$DOH ($E_{\rm up} >$ 50 K)&13 &54--194 & 0.3 & 177(71) & 4(1) $\times$ 10$^{17}$ & 7.1(2.4) $\times$ 10$^{-3}$\\

$^{\rm 13}$CH$_{\rm 3}$OH ($E_{\rm up}>$ 50 K)& 11 & 56--175 & 0.3 &  91(13) & 20(4) $\times$ 10$^{16}$ & -- \\

\hline
\multicolumn{7}{c}{outflow}\\ 
\hline

H$_{\rm 2}^{\rm 13}$CO Blue wing$^{c}$ & 3 & 20--33 &10 & 12$^{c}$ & 15(5) $\times$ 10$^{11}$  &-- \\

HDCO Blue wing$^{c}$ & 4 & 27--40 & 10 &12(7) & 9(15) $\times$ 10$^{11}$  &4.0(6.3) $\times$ 10$^{-3}$\\

HDCO Red wing$^{c}$ & 4 & 27--40 & 10 & 12(8) & 6(12) $\times$ 10$^{11}$  & 2.6(5.2) $\times$ 10$^{-3}$\\

\hline
\label{tab:RD}
\end{tabular}

$^{a}$Assumed from LVG analysis results; for the outflow component we arbitrarily assumed an extended (10$\arcsec$) size.\\
$^{b}$ To calculate the D/H ratio we assumed for HDCO and D$_2$CO the same rotational temperature of H$_2^{13}$CO ($T_{\rm rot}$ = 19 K). For CH$_2$DOH we assumed the same rotational temperature of $^{13}$CH$_3$OH ($T_{\rm rot}$ = 12 K and 91 K). \\
$^{c}$ Derived using the residual intensities after subtracting the gaussian fit of the ambient component. For the H$_{2}$$^{13}$CO wings we assumed the same $T_{\rm rot}$ of the HDCO wings (see Text).\\

\end{table*}


Why does D/H decrease from Class 0 to Class I protostars? 
Formaldehyde and methanol observed around embedded protostars
have been mostly formed at the surface of interstellar grains and 
have been then evaporated thermally when the temperature 
exceeds their temperature of sublimation. 
Two possibilities can therefore be suggested. 
The decrease of D/H from Class 0 to Class I could be due to (1) warm gas phase chemistry after the evaporation of formaldehyde 
or methanol; (2) a lower deuteration of icy formaldehyde and methanol in Class I than in Class 0. 

{\it Case 1:} Warm gas phase chemistry can decrease the deuterium fractionation
of formaldehyde and methanol through ion-neutral reactions.
Charnley et al. (1997) showed that the CH$_{2}$DOH/CH$_{3}$OH
decreases dramatically by two orders of magnitude at times longer than 3 $\times$ 10$^{5}$ yr,
because of electronic recombinations that destroy more efficiently CH$_{2}$DOH than CH$_{3}$OH.
 The timescale of 3 $\times$ 10$^{5}$ yr is consistent with the typical lifetime of Class I protostars
 (0.2-0.5 Myr; Evans et al. 2009). However, there are two problems with this picture: (i)
revised models by Osamura et al. (2004) suggests longer timescales (up to 10$^{6}$ yr), and
(ii)   the dynamical timescale of the material in the hot corino envelope
  inside the centrifigual radius could be lower (1 $\times$ 10$^{4}$-1 $\times$ 10$^{5}$ yr; Visser et al. 2009).
  In addition, the decrease of methanol deuteration occurs when
  most of the methanol is already destroyed with abundances
   lower than 1 $\times$ 10$^{-8}$. Although formaldehyde spends more time
   in the warm gas due its lower binding energy, it does not show
   any significant decrease of its deuteration (Charnley et al. 1997, Roberts \& Millar 2007).
 
\begin{figure}
\begin{center}
\includegraphics[angle=90,width=8.5cm]{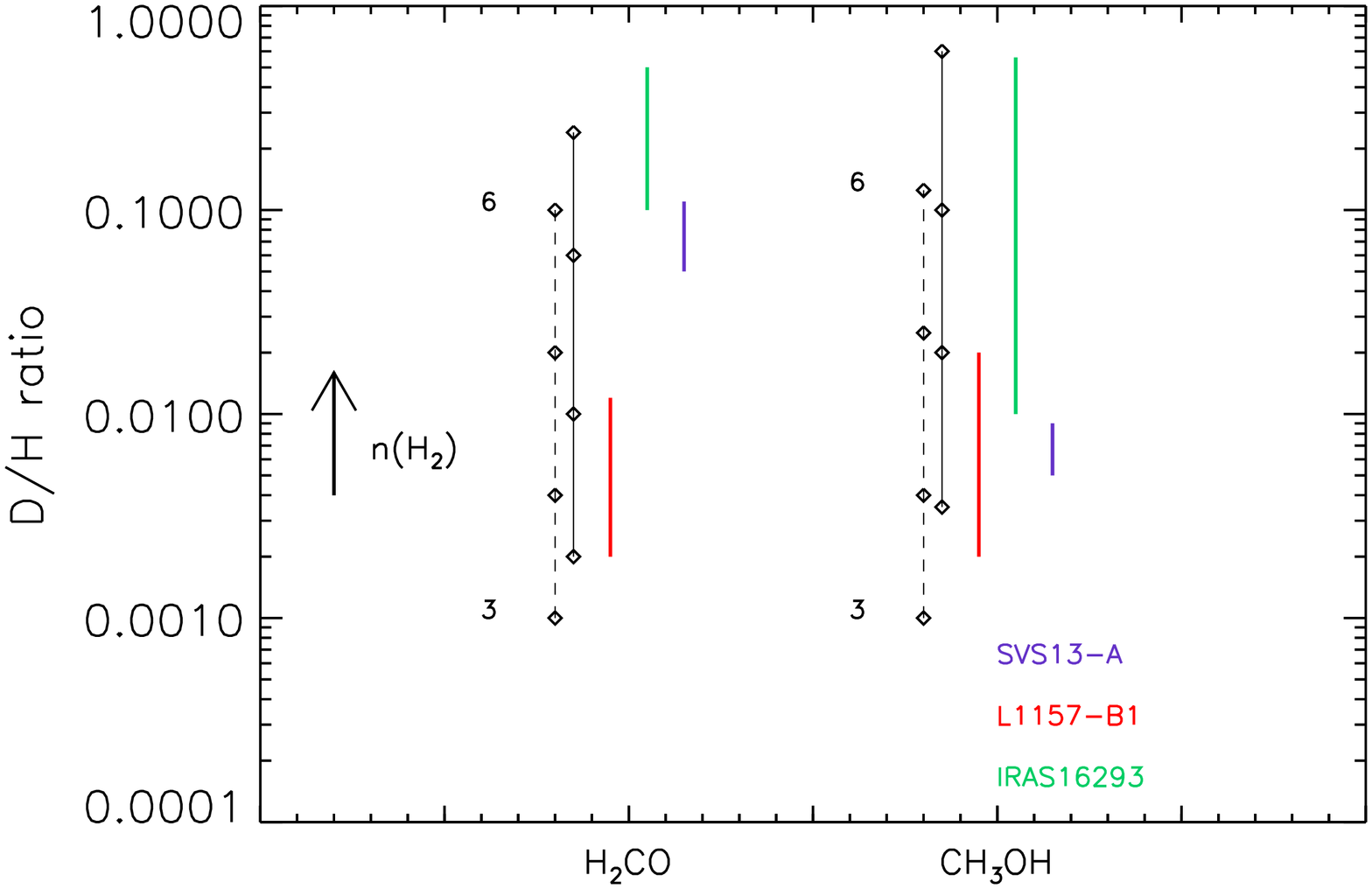}
\end{center}
\caption{Deuterium fractionation of H$_{2}$CO and CH$_{3}$OH as
found toward SVS13-A (blue), L1157-B1 (red) and IRAS16293Ð2422 (green; from Loinard et al.
2001; Parise et al. 2002, 2004). Black diamonds stand for the prediction of
Taquet et al. (2012b) for pre-shock gas densities of 10$^{3}$, 10$^{4}$, 10$^{5}$, and 10$^{6}$ cm$^{-3}$
(see labels) and temperatures of 10 (dashed line) and 20 K (solid).} 
\label{fig:Codella}
\end{figure}

{\it Case 2:}  A second possibility is that the decrease of deuteration
is due to the gradual collapse of the external shells
of the protostellar envelope. The deuterium chemistry is very
sensitive to physical (density, temperature) and chemical (CO abundance,
H$_{2}$ ortho/para ratio) parameters (see Flower et al. 2006).
Icy formaldehyde and methanol deuterations increase with the density
and with the decreasing temperature during the formation of prestellar cores
(see Taquet et al. 2012b, 2013). Taquet et al. (2014) therefore showed
that the deuteration of formaldehyde and methanol ices can decrease
by two orders of magnitude from the centre to the external part
of prestellar cores, the exact values depending on the structure
of the core and its history. In the subsequent protostellar phase,
the shells are then gradually accreted from the center to the
outer part in an inside-out fashion during the core collapse.
The methanol deuteration observed in the early Class 0 phase
would reflect the material at the center of the prestellar core
whereas the older Class I phase reflects the material coming from
the external core shells.
An instructive view has been reported by Codella et al. (2012), who
analysed the H$_2$CO and CH$_3$OH deuteration in the shocked region L1157-B1,
located relatively far (0.08 pc) from the protostar driving the shocks in the outflow, and thus
sampling an outer region probably associated with a (pre-stellar) density lower than that
where the protostar is successively born.
The D/H derived for L1157-B1 are indeed lower than what found for the standard hot corino 
IRAS16293-2422, i.e. the inner 100 AU of the protostellar core.  
In other words, H$_{2}$CO and CH$_{3}$OH deuteration can be used to measure the density
at the moment of the ices formation, before the start of the star forming process: the higher the D/H, the higher the density.
In the case of SVS13-A the D/H for formaldehyde in the outflow,
sampling a region definitely more extended than the protostellar
high density cocoon, is indeed supporting this scenario (see Figure \ref{fig:Codella}).
The decrease by two orders of magnitude from Class 0 to Class I protostars 
observed for the D$_{2}$CO/H$_{2}$CO and CH$_{2}$DOH/CH$_{3}$OH ratios
is in good agreement with the model predictions by Taquet et al. (2014) 
within an order of magnitude
although the models still tend to underpredict the absolute ratios.
It should be noted that the decrease of formaldehyde and methanol
deuterations with the evolutionary stage of the protostar are not 
necessarily accompanied by a decrease of water deuteration. 
As water ice is mostly formed in molecular clouds before the formation of 
prestellar cores, its deuteration only weakly varies within prestellar cores. 
This scenario can therefore simultaneously explain the decrease of deuteration 
of formaldehyde and methanol observed in this work and, in addition, the constant deuteration
of water observed towards SVS13-A by Codella et al. (2016).

\section{Conclusions}

We studied the formaldehyde and methanol deuteration in the Class I object
SVS13-A with the IRAM 30-m antenna in the framework 
of the ASAI large program consisting of an unbiased spectral survey at 1.3, 2 and 3 mm
towards the source.
The aim of this project was to understand how the deuterium fractionation of organics like
H$_{2}$CO and CH$_{3}$OH change in a Class I object, SVS13-A, with respect to the Class 0 sources.
The bulk of the detected lines are in the 1.3 mm band corresponding 
to a telescope HPBW $\sim$ 10$\arcsec$.
This ensures than the signal is coming from SVS13-A and it is not contaminated
by the SVS13-B Class 0 object, offset by 15$\arcsec$. 
The main results are reported as follows:

\begin{enumerate}

\item
We detected 7 lines of H$_{2}$$^{13}$CO, 5 transitions of HDCO and 5 lines of D$_{2}$CO
with excitation energies $E_{\rm up}$ in the 10-45 K range. 
The LVG analysis of H$_{2}$$^{13}$CO indicates low values of $T_{\rm kin}$ ($\sim$ 20 K),
densities larger than 10$^{6}$ cm$^{-3}$, and an emitting size of about 5$\arcsec$ ($\sim$ 1200 AU).
The low temperature is confirmed by the rotational diagram performed for all formaldehyde 
isotopologues, suggesting the association with the molecular envelope surrounding the protostar.

\item
Both H$_{2}$$^{13}$CO and HDCO lines show wings indicating emission from outflowing gas.
For both the blue- and the red-shifted emission we obtained a low $T_{\rm rot}$ ($\sim$ 12 K),
in agreement with the association with the extended outflow driven by SVS13-A.

\item
We detected 18 lines of $^{13}$CH$_{3}$OH and 27 transitions of CH$_{2}$DOH with 
$E_{\rm up}$ in the 20-276 K range.
We report the detection of CHD$_{2}$OH and CH$_{3}$OD through 2 different
transition for each species. The LVG analysis of $^{13}$CH$_{3}$OH suggests the occurrence of of two components,
with different excitation conditions:
(1) a compact region ($\theta_{s}$ $\simeq$ 0$\farcs$3, 70 AU) corresponding to
high temperatures ($T_{\rm kin} \sim$ 80 K) and very high densities ($>$ 10$^{8}$ cm$^{-3}$),
clearly being the hot-corino (recently discovered by HDO observations; Codella et al. 2016). 
(2) a colder ($T_{\rm kin} \leq$ 70 K), more extended ($\theta_{s}$ $\simeq$ 2$\arcsec$ -- 4$\arcsec$)
region associated with densities $>$ 10$^{6}$ cm$^{-3}$.
The rotation diagram analysis confirms for the deuterated methanol 
the presence of a hot corino component associated
to high densities and temperatures and a second component
due to colder gas emission.

\item
We measured for formaldehyde D/H $\sim$ 9 $\times$ 10$^{-2}$,
a value consistent with the average value reported from Class 0 sources (D/H $\sim$ 0.12, Parise et al. 2006).
The deuterium fractionation derived for the outflowing component
is D/H $\sim$ 4 $\times$ 10$^{-3}$, in agreement with those measured
 in the shocked region associated with the L1157 protostellar outflow
by Codella et al. (2012).
On the other hand, for D$_{2}$CO we obtained D/H $\sim$ 4 $\times$ 10$^{-3}$, lower by one order of magnitude with respect to Class 0 objects.
This trend is even stronger for the measured methanol deuteration, which is 4 $\times$ 10$^{-3}$,
two orders of magnitude lower than the values reported by Parise et al. (2006)
for Class 0 objects.

\item
The detection of CH$_{3}$OD allows us to derive a measure of
the [CH$_{2}$DOH]/[CH$_{3}$OD] ratio that is in the 2.0 Ð- 2.5 range.
Previous measurements by Ratajczak et al. (2011), including
both low- and high-mass star forming regions, indicate a weak trend with
a lower abundance ratio observed in massive hot cores with respect to
 (low-mass) hot- corinos (as well as in intermediate-mass
protostars), by typically one order of magnitude.
According to these indications, in SVS13-A CH$_{3}$OD was expected to be less abundant than
CH$_{2}$DOH, by more than a factor 10 (Ratajczak et al. 2011).
However our [CH$_{2}$DOH]/[CH$_{3}$OD] measurement questions 
the previous indication about a correlation between this ratio and the protostellar luminosity.

\item
The low deuterium fractionation measured towards SVS13-A
could be an indication of the modified chemical content in the evolutionary transition
from the Class 0 phase to the Class I phase.
Alternatively, the decrease of D/H in a more evolved phase could be due to the gradual collapse
of the external shells of the protostellar evelope, less deuterated because composed of 
ices formed in a less dense region.
Only high resolution interferometric observations, able to sample 
the inner region of the protostar ($<$ 1$\arcsec$ corresponding to $\sim$ 235 AU
at the source distance) and to disentangle the emission coming from the different
 protostar components, will properly answer these open questions.
 
 \end{enumerate}

\section*{Acknowledgments}

The authors are grateful 
to the IRAM staff for its help in the
calibration of the 30-m data. The research leading to these results
has received funding from the European Commission Seventh Framework
Programme (FP/2007-2013) under grant agreement N¡ 283393 (RadioNet3).
This work was partly supported by the PRIN INAF 2012 -- JEDI and by
the Italian Ministero dell'Istruzione, Universit\`a e Ricerca through
the grant Progetti Premiali 2012 -- iALMA wich is also founding 
the EB PhD project.
BL and CCe acknowledge the
financial support from the French Space Agency CNES
and RB from Spanish MINECO (through project FIS2012-32096).
BL and CCe acknowledge support from the CNRS program 
ÓPhysique et Chimie du Milieu InterstellaireÓ
(PCMI) and a grant from LabeX Osug@2020(Investissements dÕavenir - ANR10LABX56).

{}

\bsp

\label{lastpage}

\end{document}